\documentclass[12pt]{article}
\usepackage{graphicx}
\usepackage{epsf} 
\def\beq{\begin{equation}}
\def\eeq{\end{equation}}
\def\bea{\begin{eqnarray}}
\def\eea{\end{eqnarray}}
\def\nn{\nonumber}
\def\sss{\scriptscriptstyle}
\def\roughly#1{\mathrel{\raise.3ex\hbox
{$#1$\kern-.75em\lower1ex\hbox{$\sim$}}}}

\def\bd{B_d^0}

\def\btod{b \to d}
\def\btos{b \to s}
\def\ks{K_{\sss S}}

\def\pew{P_{\sss EW}}
\def\pewc{P^{\sss C}_{\sss EW}}
\def\pewpu{P^{\sss P_u}_{\sss EW}}
\def\pewe{P^{\sss E}_{\sss EW}}
\def\pewa{P^{\sss A}_{\sss EW}}
\def\pewpau{P^{\sss PA_u}_{\sss EW}}
\def\pewp{P'_{\sss EW}}
\def\pewcp{P^{\prime \sss C}_{\sss EW}}
\def\pewpup{P^{\prime \sss P_u}_{\sss EW}}
\def\pewep{P^{\prime \sss E}_{\sss EW}}
\def\pewap{P^{\prime \sss A}_{\sss EW}}
\def\pewpaup{P^{\prime \sss PA_u}_{\sss EW}}

\def\btopipi{B \to \pi \pi}
\def\btopik{B \to \pi K}
\def\bra#1{\left\langle #1\right|}
\def\ket#1{\left| #1\right\rangle}
\def\gf{G_{\sss F}}
\def\wt{\widetilde}
\makeatletter
\newcommand{\contraction}[5][1ex]{%
 \mathchoice
  {\contraction@\displaystyle{#2}{#3}{#4}{#5}{#1}}%
  {\contraction@\textstyle{#2}{#3}{#4}{#5}{#1}}%
  {\contraction@\scriptstyle{#2}{#3}{#4}{#5}{#1}}%
  {\contraction@\scriptscriptstyle{#2}{#3}{#4}{#5}{#1}}}%
\newcommand{\contraction@}[6]{%
 \setbox0=\hbox{$#1#2$}%
 \setbox2=\hbox{$#1#3$}%
 \setbox4=\hbox{$#1#4$}%
 \setbox6=\hbox{$#1#5$}%
 \dimen0=\wd2%
 \advance\dimen0 by \wd6%
 \divide\dimen0 by 2%
 \advance\dimen0 by \wd4%
 \vbox{%
  \hbox to 0pt{%
   \kern \wd0%
   \kern 0.4\wd2%pour deplacer horizontalement!!!!!
   \contraction@@{\dimen0}{#6}%
   \hss}%
  \vskip 0.8ex%pour deplacer verticalement!!!!!
  \vskip\ht2}}

\newcommand{\contraction@@}[3][0.06em]{%
 \hbox{%
  \vrule width #1 height 0pt depth #3%
  \vrule width #2 height 0pt depth #1%
  \vrule width #1 height 0pt depth #3%
  \relax}}
\makeatother

\begin{document}

\begin{flushright}
UdeM-GPP-TH-06-145 \\
\end{flushright}

\begin{center}
\bigskip
{\Large \bf \boldmath Hadronic $B$ Decays: A General Approach} \\
\bigskip
\bigskip
Maxime Imbeault $^{a,}$\footnote{maxime.imbeault@umontreal.ca},
Alakabha Datta $^{b,}$\footnote{datta@physics.utoronto.ca},
and David London $^{a,}$\footnote{london@lps.umontreal.ca}
\end{center}
\begin{flushleft}
~~~~~~~~~~~$a$: {\it Physique des Particules, Universit\'e
de Montr\'eal,}\\
~~~~~~~~~~~~~~~{\it C.P. 6128, succ. centre-ville, Montr\'eal, QC,
Canada H3C 3J7}\\
~~~~~~~~~~~$b$: {\it Department of Physics, University of Toronto,}\\
~~~~~~~~~~~~~~~{\it 60 St.\ George Street, Toronto, ON, Canada M5S
1A7}\\
\end{flushleft}

\begin{center}
\bigskip (\today)
\vskip0.5cm {\Large Abstract\\} \vskip3truemm
\parbox[t]{\textwidth}{In this paper, we propose a general approach
for describing hadronic $B$ decays. Using this method, all amplitudes
for such decays can be expressed in terms of contractions, though the
matrix elements are not evaluated. Many years ago, Buras and
Silvestrini proposed a similar approach. However, our technique goes
beyond theirs in several ways. First, we include recent theoretical
and experimental developments which indicate which contractions are
negligible, and which are expected to be smaller than others. Second,
we show that all $B$-decay diagrams can be simply expressed in terms
of contractions. This constitutes a formal proof that the diagrammatic
method is rigourous. Third, we show that one reproduces the relations
between tree and electroweak-penguin diagrams described by Neubert and
Rosner, and by Gronau, Pirjol and Yan. Fourth, although the previous
results hold to all orders in $\alpha_s$, we show that it is also
possible to work order-by-order in this approach. In this way it is
possible to make a connection with the matrix-element evaluation
methods of QCD factorization (QCDfac) and perturbative QCD (pQCD).
Finally, using the contractions approach, we re-evaluate the question
of whether there is a ``$\btopik$ puzzle.'' At $O(\alpha_s^0)$, we
find that the diagram ratio $|C'/T'|$ is about 0.17, a factor of 10
too small to explain all the $\btopik$ data. Both QCDfac and pQCD find
that, at $O(\alpha_s^1)$, the value of $|C'/T'|$ may be raised to only
about 2-3 times its lowest-order value. We therefore conclude that,
assuming the effect is not a statistical fluctuation, it is likely
that the value of $|C'/T'|$ is similar to its $O(\alpha_s^0)$ result,
and that there really is a $\btopik$ puzzle.}
\end{center}

\thispagestyle{empty}
\newpage
\setcounter{page}{1}
% Decrease texheight (for preprint numbers) again
%\textheight 23.0 true cm
\baselineskip=14pt

\section{Introduction}

The study of hadronic $B$ decays\footnote{In this paper, we present
results for ${\bar B}$ mesons, which contain a $b$ quark. We will
nevertheless continue to refer to $B$ decays.} has a long and storied
history. Much work has gone into all aspects of this problem.

One important development was the operator-product expansion (OPE).
Here the effective Hamiltonian for quark-level $b$ decays takes the
form \cite{BBL}
\beq 
H_{\sss eff} = {\gf \over \sqrt{2}} \sum_{q=d,s}
\left(\sum_{p=u,c}\lambda_p^{(q)} (C_1(\mu) O^p_1 (\mu) + C_2(\mu)
O^p_2 (\mu)) - \lambda_t^{(q)} \sum_{i=3}^{10} C_i(\mu) O_i(\mu)
\right) ~,
\label{Heff}
\eeq
where $\lambda_p^{(q)}=V_{pb} V^*_{pq}$. $\mu$ is the renormalization
point, typically taken to be $O(m_b)$. The Wilson coefficients $C_i$
(WC's) include gluons (QCD corrections) whose energy is above $\mu$
(short distance), while the operators $O_i$ include QCD corrections of
energy less than $\mu$ (long distance). All physical quantities must
be independent of $\mu$. Note: factors of $G_F/\sqrt{2}$ are omitted
for the remainder of this paper.

The operators take the following form:
\beq
  O_1^p = (\bar p_\alpha b_\alpha)_{V-A}\,
  (\bar q_\beta p_\beta)_{V-A} ~~,~~~~
  O_2^p = (\bar p_\alpha b_\beta)_{V-A}\,
  (\bar q_\beta p_\alpha)_{V-A} ~,
\eeq
summed over colour indices $\alpha$ and $\beta$. These are the usual
(tree-level) current--current operators induced by $W$-boson exchange.
\bea
  O_3 = (\bar q_\alpha b_\alpha)_{V-A}\,\sum_{q'}\,
  (\bar q'_\beta q'_\beta)_{V-A} & , &
  O_4 = (\bar q_\alpha b_\beta)_{V-A}\,\sum_{q'}\,
  (\bar q'_\beta q'_\alpha)_{V-A} \,, \nonumber\\
  O_5 = (\bar q_\alpha b_\alpha)_{V-A}\,\sum_{q'}\,
  (\bar q'_\beta q'_\beta)_{V+A} & , &
  O_6 = (\bar q_\alpha b_\beta)_{V-A}\,\sum_{q'}\,
  (\bar q'_\beta q'_\alpha)_{V+A} \,,
\eea
summed over the light flavors $q'=u,d,s,c$. These are referred to as
QCD penguin operators.
\bea
  O_7 = \frac32\,(\bar q_\alpha b_\alpha)_{V-A}\,\sum_{q'}\,e_{q'}\,
  (\bar q'_\beta q'_\beta)_{V+A} & , &
  O_8 = \frac32\,(\bar q_\alpha b_\beta)_{V-A}\,\sum_{q'}\,e_{q'}\,
  (\bar q'_\beta q'_\alpha)_{V+A} \,, \nonumber\\
  O_9 = \frac32\,(\bar q_\alpha b_\alpha)_{V-A}\,\sum_{q'}\,e_{q'}\,
  (\bar q'_\beta q'_\beta)_{V-A} & , &
  O_{10} = \frac32\,(\bar q_\alpha b_\beta)_{V-A}\,\sum_{q'}\,e_{q'}\,
  (\bar q'_\beta q'_\alpha)_{V-A} \,,
\eea
with $e_{q'}$ denoting the electric charges of the quarks. These are
called electroweak-penguin (EWP) operators. The quark current $(\bar
q_1 q_2)_{V\pm A}$ denotes $\bar q_1\gamma^\mu (1\pm\gamma_5)q_2$. All
quark-level $b$ decays can be expressed in terms of the effective
Hamiltonian.

An alternative description is the diagrammatic approach \cite{GHLR},
in which quark-level $b$ decays are characterized by various
topologies. These include the colour-favored and colour-suppressed
tree amplitudes $T$ and $C$, the gluonic penguin amplitude $P$, the
colour-favored and colour-suppressed EWP amplitudes $\pew$ and
$\pewc$, the exchange diagram $E$, the annihilation amplitude $A$, and
the penguin-annihilation diagram $PA$. This formalism has been
extensively used, but the relation to the OPE and the effective
Hamiltonian has not been made entirely clear.

Of course, the true decays involve mesons, and do not take place at
the quark level. That is, the decay is $B\to F$, where $B$ is a
charged or neutral $B$ meson, and the final state $F$ involves mesonic
states. Thus, one has to calculate $\langle F | H_{eff} | B \rangle$,
which involves the hadronic matrix elements $\langle F | O_i(\mu) | B
\rangle$. There are basically two competing approaches to calculating
matrix elements: QCD factorization (QCDfac) \cite{BBNS} and
perturbative QCD (pQCD) \cite{PQCD}.

The various WC's have been calculated at next-to-leading order
\cite{BBL}. It is found that $C_7$ and $C_8$ are negligible, so that,
to a good approximation, the EWPs are purely $(V-A)\times(V-A)$. The
consequences of this were investigated by Neubert and Rosner
\cite{NR}, and by Gronau, Pirjol and Yan (GPY) \cite{GPY}. Using this
approximation and flavour symmetries, GPY found two relationships
between the EWP diagrams and the other $(V-A)\times(V-A)$ diagrams
($T$, $C$, $E$, $A$) in $\btopipi$ and $\btopik$ decays.

{}From this brief summary, we see that there are many different
aspects to hadronic $B$ decays. In the present paper, we propose an
approach which makes a connection to all of these. It is based on
contractions. When calculating the amplitude for a particular decay,
one must ``sandwich'' all operators of the effective Hamiltonian
between initial and final states. All such terms have the form
\beq
\langle \bar q_1 q_2 \bar q_3 q_4 | \bar q_5 b \, \bar q_6 q_7 |
\bar q_8 b \rangle~.
\label{contracform}
\eeq
(Dirac and colour structures are omitted for notational convenience.)
Here $\bar q_1 q_2 \bar q_3 q_4$ are the final-state quarks that
eventually hadronize into the final-state mesons via the strong
interactions. In our method, we are interested in the possible ways in
which the final-state quarks can be produced in $B$ decays through the
effective Hamiltonian. These can be obtained by simply applying the
basic rules of quantum field theory, and summing over all possible
Wick contractions of all operators.

Even though specific contractions between pairs of quarks in the decay
process are involved, it is understood that any number of gluons may
be exchanged between these quarks. Indeed, such exchanges are
necessary in order for the final-state quarks to hadronize into
mesons. The results of our general approach therefore hold to all
orders in $\alpha_s$. In our method we are not interested in the
process of hadronization of the final-state quarks into mesons. If one
wishes to calculate this, it is necessary to resort to models like
QCDfac and pQCD.

We will show that one can simply express all the diagrams in
Ref.~\cite{GHLR} in terms of contractions. Since the set of all
possible contractions follows directly from the effective Hamiltonian,
this demonstrates explicitly the connection between diagrams and the
OPE. As such, it constitutes a formal proof that the diagrammatic
approach is rigourous. We also show that one reproduces the relations
of GPY. Furthermore, we show that we can include gluons in the
contractions order-by-order. (Of course, for such an exercise to have
any meaning, one has to argue that it is possible to calculate
nonleptonic decays by including gluon exchanges order-by-order. The
frameworks of QCDfac, pQCD and soft collinear effective theory (SCET)
\cite{SCET} have all discussed when and where an order-by-order
expansion in $\alpha_s$ of hadronic decays is possible.)

Contractions were analyzed some years ago by Buras and Silvestrini
(BS) \cite{BS}. At the most basic level, our method is simply a
reorganization of BS. However, the connection between diagrams and
contractions is much simpler in our approach. In addition, since the
appearance of Ref.~\cite{BS}, we have obtained a variety of insights
into nonleptonic decays. For example, there have been important
developments in the various theoretical approaches. As such, the
contractions can be separated into various classes, and the
contractions in certain classes can be argued to be small because they
arise only at $O(1/m_b)$. In addition, numerous measurements in $B$
decays have been made, and these help us to determine which
contractions may be important. For instance, we know that certain
rescattering contractions are in general smaller than other
non-rescattering contractions. (We describe this in more detail in
Sec.~2.1.)

Using this general approach, we can express the amplitude for any
decay fully in terms of contractions (indeed, this is what BS have
done). However, as discussed above, some of these contractions can be
argued to be negligible, either for theoretical or experimental
reasons. We can therefore simplify the expressions with only minimal
theory or model inputs. A fit to experimental data then allows us to
obtain the various contractions. The fitted values can then be used to
compare with theory calculations or make predictions for other
processes. Note that our approach is valid not only for the SM but
also for new-physics (NP) contributions. Hence, knowledge of the
contractions from a fit to the data can give information about the
nature of NP \cite{Datta-Lon}.

In Sec.~2, we present the contractions method. The relation with the
approach of BS is shown here. We make the connection to diagrams in
Sec.~3. We show that one reproduces the GPY relations in Sec.~4, by
working to all orders in $\alpha_s$. If one works only order-by-order
in $\alpha_s$, one can make a connection between contractions and the
matrix-element formalisms of QCDfac and pQCD. This is shown in
Sec.~5. In Sec.~6, we examine whether there really is a ``$\btopik$
puzzle.'' We conclude in Sec.~7.

\section{Contractions}

\subsection{General Considerations}

We begin by describing the contractions in general. For a given $B$
decay, one obtains many matrix elements of the form $\langle \bar q_1
q_2 \bar q_3 q_4 | \bar q_5 b \, \bar q_6 q_7 | \bar q_8 b \rangle$
[Eq.~(\ref{contracform})], where Dirac and colour structures have
been suppressed. $\bar q_8 b$ is the $B$ meson. This matrix element
describes the decay $B \to M_1 M_2$, where $M_{1,2}$ are mesons. The
final-state mesons contain the quarks $\bar q_1$, $q_2$, $\bar q_3$
and $q_4$, but we are free to choose the quark assignments as we
wish. In the following, we choose $M_1 = \bar q_1 q_2$ and $M_2 =\bar
q_3 q_4$. In addition, we will write the final state for such $B$
decays as
\beq
{1 \over \sqrt{2}} \left[ M_1 M_2 + M_2 M_1 \right] ~.
\label{decayM1M2}
\eeq
The reason we do this is as follows. In $B\to\pi\pi$ decays, isospin
requires that we symmetrize the final-state $\pi$'s. We choose the
above form for the $M_1 M_2$ final state in order to resemble
$B\to\pi\pi$.  However, we stress that the analysis in this paper
would not change with another choice, either of the $M_{1,2}$ quark
assignments, or of the form of the $M_1 M_2$ final state. For example,
in $\btopik$ we can choose $M_1 = \bar q_1 q_2= \pi$ and $M_2 =\bar
q_3 q_4= K$, or $M_1 = \bar q_1 q_2=K$ and $M_2 =\bar q_3 q_4=\pi$
without changing the physics. We refer to the freedom to exchange
$\bar q_1 q_2 \leftrightarrow \bar q_3 q_4$ as {\it final-state
symmetry.}

For a given $b$ decay, there are 24 possible contractions:
\bea
{\underline A}=
{
\contraction[1ex]{\langle \bar q_1 }{q_2 }{}{\bar q_3 }
\contraction[1ex]{\langle \bar q_1 q_2 \bar q_3 }{q_4 }{| \bar q_5 b }{\bar q_6 }
\contraction[2ex]{\langle \bar q_1 q_2 \bar q_3 q_4 | }{\bar q_5 }{b \bar q_6 }{q_7 }
\contraction[3ex]{\langle }{\bar q_1 }{q_2 \bar q_3 q_4 | \bar q_5 b \bar q_6 q_7 | }{\bar q_8}
\langle \bar q_1 q_2 \bar q_3 q_4 | \bar q_5 b \bar q_6 q_7 | \bar q_8 b \rangle
}~,
&~~
{\underline B}=
{
\contraction[1ex]{\langle \bar q_1 }{q_2 }{}{\bar q_3 }
\contraction[1ex]{\langle \bar q_1 q_2 \bar q_3 }{q_4 }{| \bar q_5 b }{\bar q_6 }
\contraction[2ex]{\langle }{\bar q_1 }{ q_2 \bar q_3 q_4 | \bar q_5 b \bar q_6 }{q_7 }
\contraction[3ex]{\langle \bar q_1 q_2 \bar q_3 q_4 | }{\bar q_5 }{b \bar q_6 q_7 | }{\bar q_8 }
\langle \bar q_1 q_2 \bar q_3 q_4 | \bar q_5 b \bar q_6 q_7 | \bar q_8 b \rangle
}~,
&~~
{\underline C}=
{
\contraction[1ex]{\langle }{\bar q_1 }{}{q_2 }
\contraction[1ex]{\langle \bar q_1 q_2 \bar q_3 q_4 | }{\bar q_5 }{b \bar q_6 }{q_7 }
\contraction[2ex]{\langle \bar q_1 q_2 \bar q_3 }{q_4 }{| \bar q_5 b }{\bar q_6 }
\contraction[3ex]{\langle \bar q_1 q_2 }{\bar q_3 }{q_4 | \bar q_5 b \bar q_6 q_7 | }{\bar q_8 }
\langle \bar q_1 q_2 \bar q_3 q_4 | \bar q_5 b \bar q_6 q_7 | \bar q_8 b \rangle
}~,
\nn\\
{\underline D}=
{
\contraction[1ex]{\langle }{\bar q_1 }{}{q_2 }
\contraction[1ex]{\langle \bar q_1 q_2 \bar q_3 }{q_4 }{| \bar q_5 b }{\bar q_6 }
\contraction[2ex]{\langle \bar q_1 q_2 \bar q_3 q_4 | }{\bar q_5 }{b \bar q_6 q_7 | }{\bar q_8 }
\contraction[3ex]{\langle \bar q_1 q_2 }{\bar q_3 }{q_4 | \bar q_5 b \bar q_6 }{q_7 }
\langle \bar q_1 q_2 \bar q_3 q_4 | \bar q_5 b \bar q_6 q_7 | \bar q_8 b \rangle
}~,
&~~
{\underline E}=
{
\contraction[1ex]{\langle \bar q_1 q_2 \bar q_3 }{q_4 }{| \bar q_5 b }{\bar q_6 }
\contraction[2ex]{\langle \bar q_1 q_2 }{\bar q_3 }{q_4 | \bar q_5 b \bar q_6 }{q_7 }
\contraction[3ex]{\langle \bar q_1 }{q_2 }{\bar q_3 q_4 | }{\bar q_5 }
\contraction[4ex]{\langle }{\bar q_1 }{q_2 \bar q_3 q_4 | \bar q_5 b \bar q_6 q_7 | }{\bar q_8}
\langle \bar q_1 q_2 \bar q_3 q_4 | \bar q_5 b \bar q_6 q_7 | \bar q_8 b \rangle
}~,
&~~
{\underline F}=
{
\contraction[1ex]{\langle \bar q_1 }{q_2 }{\bar q_3 q_4 | }{\bar q_5 }
\contraction[2ex]{\langle \bar q_1 q_2 \bar q_3 }{q_4 }{| \bar q_5 b }{\bar q_6 }
\contraction[3ex]{\langle \bar q_1 q_2 }{\bar q_3 }{q_4 | \bar q_5 b \bar q_6 q_7 | }{\bar q_8 }
\contraction[4ex]{\langle }{\bar q_1 }{ q_2 \bar q_3 q_4 | \bar q_5 b \bar q_6 }{q_7 }
\langle \bar q_1 q_2 \bar q_3 q_4 | \bar q_5 b \bar q_6 q_7 | \bar q_8 b \rangle
}~,
\nn\\
{\underline G}=
{
\contraction[1ex]{\langle \bar q_1 }{q_2 }{}{\bar q_3 }
\contraction[1ex]{\langle \bar q_1 q_2 \bar q_3 }{q_4 }{| }{\bar q_5 }
\contraction[1ex]{\langle \bar q_1 q_2 \bar q_3 q_4 | \bar q_5 b }{\bar q_6 }{}{q_7}
\contraction[2ex]{\langle }{\bar q_1 }{q_2 \bar q_3 q_4 | \bar q_5 b \bar q_6 q_7 | }{\bar q_8}
\langle \bar q_1 q_2 \bar q_3 q_4 | \bar q_5 b \bar q_6 q_7 | \bar q_8 b \rangle
}~,
&~~
{\underline H}=
{
\contraction[1ex]{\langle \bar q_1 }{q_2 }{}{\bar q_3 }
\contraction[1ex]{\langle \bar q_1 q_2 \bar q_3 }{q_4 }{| }{\bar q_5 }
\contraction[1ex]{\langle \bar q_1 q_2 \bar q_3 q_4 | \bar q_5 b }{\bar q_6 }{q_7 | }{\bar q_8 }
\contraction[2ex]{\langle }{\bar q_1 }{ q_2 \bar q_3 q_4 | \bar q_5 b \bar q_6 }{q_7 }
\langle \bar q_1 q_2 \bar q_3 q_4 | \bar q_5 b \bar q_6 q_7 | \bar q_8 b \rangle
}~,
&~~
{\underline I}=
{
\contraction[1ex]{\langle \bar q_1 q_2 \bar q_3 }{q_4 }{| }{\bar q_5 }
\contraction[2ex]{\langle \bar q_1 q_2 }{\bar q_3 }{q_4 | \bar q_5 b \bar q_6 }{q_7 }
\contraction[3ex]{\langle \bar q_1 }{q_2 }{\bar q_3 q_4 | \bar q_5 b }{\bar q_6 }
\contraction[4ex]{\langle }{\bar q_1 }{q_2 \bar q_3 q_4 | \bar q_5 b \bar q_6 q_7 | }{\bar q_8}
\langle \bar q_1 q_2 \bar q_3 q_4 | \bar q_5 b \bar q_6 q_7 | \bar q_8 b \rangle
}~,
\nn\\
{\underline J}=
{
\contraction[1ex]{\langle }{\bar q_1 }{}{q_2 }
\contraction[1ex]{\langle \bar q_1 q_2 }{\bar q_3 }{}{q_4 }
\contraction[2ex]{\langle \bar q_1 q_2 \bar q_3 q_4 | }{\bar q_5 }{b \bar q_6 q_7 | }{\bar q_8 }
\contraction[1ex]{\langle \bar q_1 q_2 \bar q_3 q_4 | \bar q_5 b }{\bar q_6 }{}{q_7}
\langle \bar q_1 q_2 \bar q_3 q_4 | \bar q_5 b \bar q_6 q_7 | \bar q_8 b \rangle
}~,
&~~
{\underline K}=
{
\contraction[1ex]{\langle }{\bar q_1 }{}{q_2 }
\contraction[1ex]{\langle \bar q_1 q_2 \bar q_3 }{q_4 }{| }{\bar q_5 }
\contraction[1ex]{\langle \bar q_1 q_2 \bar q_3 q_4 | \bar q_5 b }{\bar q_6 }{}{q_7}
\contraction[2ex]{\langle \bar q_1 q_2 }{\bar q_3 }{q_4 | \bar q_5 b \bar q_6 q_7 | }{\bar q_8 }
\langle \bar q_1 q_2 \bar q_3 q_4 | \bar q_5 b \bar q_6 q_7 | \bar q_8 b \rangle
}~,
&~~
{\underline L}=
{
\contraction[1ex]{\langle }{\bar q_1 }{}{q_2 }
\contraction[1ex]{\langle \bar q_1 q_2 \bar q_3 }{q_4 }{| }{\bar q_5 }
\contraction[1ex]{\langle \bar q_1 q_2 \bar q_3 q_4 | \bar q_5 b }{\bar q_6 }{q_7 | }{\bar q_8 }
\contraction[2ex]{\langle \bar q_1 q_2 }{\bar q_3 }{q_4 | \bar q_5 b \bar q_6 }{q_7 }
\langle \bar q_1 q_2 \bar q_3 q_4 | \bar q_5 b \bar q_6 q_7 | \bar q_8 b \rangle
}~,
\nn\\
{\underline M}=
{
\contraction[1ex]{\langle \bar q_1 q_2 \bar q_3 }{q_4 }{| }{\bar q_5 }
\contraction[2ex]{\langle \bar q_1 }{q_2 }{\bar q_3 q_4 | \bar q_5 b }{\bar q_6 }
\contraction[3ex]{\langle }{\bar q_1 }{ q_2 \bar q_3 q_4 | \bar q_5 b \bar q_6 }{q_7 }
\contraction[4ex]{\langle \bar q_1 q_2 }{\bar q_3 }{q_4 | \bar q_5 b \bar q_6 q_7 | }{\bar q_8 }
\langle \bar q_1 q_2 \bar q_3 q_4 | \bar q_5 b \bar q_6 q_7 | \bar q_8 b \rangle
}~,
&~~
{\underline N}=
{
\contraction[1ex]{\langle \bar q_1 }{q_2 }{}{\bar q_3 }
\contraction[2ex]{\langle }{\bar q_1 }{ q_2 \bar q_3 }{q_4 }
\contraction[1ex]{\langle \bar q_1 q_2 \bar q_3 q_4 | \bar q_5 b }{\bar q_6 }{}{q_7}
\contraction[2ex]{\langle \bar q_1 q_2 \bar q_3 q_4 | }{\bar q_5 }{b \bar q_6 q_7 | }{\bar q_8 }
\langle \bar q_1 q_2 \bar q_3 q_4 | \bar q_5 b \bar q_6 q_7 | \bar q_8 b \rangle
}~,
&~~
{\underline O}=
{
\contraction[1ex]{\langle \bar q_1 q_2 }{\bar q_3 }{}{q_4 }
\contraction[1ex]{\langle \bar q_1 q_2 \bar q_3 q_4 | }{\bar q_5 }{b \bar q_6 }{q_7 }
\contraction[2ex]{\langle \bar q_1 }{q_2 }{\bar q_3 q_4 | \bar q_5 b }{\bar q_6 }
\contraction[3ex]{\langle }{\bar q_1 }{q_2 \bar q_3 q_4 | \bar q_5 b \bar q_6 q_7 | }{\bar q_8}
\langle \bar q_1 q_2 \bar q_3 q_4 | \bar q_5 b \bar q_6 q_7 | \bar q_8 b \rangle
}~,
\nn\\
{\underline P}=
{
\contraction[1ex]{\langle \bar q_1 q_2 }{\bar q_3 }{}{q_4 }
\contraction[2ex]{\langle \bar q_1 }{q_2 }{\bar q_3 q_4 | \bar q_5 b }{\bar q_6 }
\contraction[3ex]{\langle }{\bar q_1 }{ q_2 \bar q_3 q_4 | \bar q_5 b \bar q_6 }{q_7 }
\contraction[1ex]{\langle \bar q_1 q_2 \bar q_3 q_4 | }{\bar q_5 }{b \bar q_6 q_7 | }{\bar q_8 }
\langle \bar q_1 q_2 \bar q_3 q_4 | \bar q_5 b \bar q_6 q_7 | \bar q_8 b \rangle
}~,
&~~
{\underline Q}=
{
\contraction[1ex]{\langle }{\bar q_1 }{ q_2 \bar q_3 }{q_4 }
\contraction[1ex]{\langle \bar q_1 q_2 \bar q_3 q_4 | }{\bar q_5 }{b \bar q_6 }{q_7 }
\contraction[2ex]{\langle \bar q_1 q_2 }{\bar q_3 }{q_4 | \bar q_5 b \bar q_6 q_7 | }{\bar q_8 }
\contraction[3ex]{\langle \bar q_1 }{q_2 }{\bar q_3 q_4 | \bar q_5 b }{\bar q_6 }
\langle \bar q_1 q_2 \bar q_3 q_4 | \bar q_5 b \bar q_6 q_7 | \bar q_8 b \rangle
}~,
&~~
{\underline R}=
{
\contraction[1ex]{\langle }{\bar q_1 }{ q_2 \bar q_3 }{q_4 }
\contraction[2ex]{\langle \bar q_1 }{q_2 }{\bar q_3 q_4 | \bar q_5 b }{\bar q_6 }
\contraction[3ex]{\langle \bar q_1 q_2 }{\bar q_3 }{q_4 | \bar q_5 b \bar q_6 }{q_7 }
\contraction[1ex]{\langle \bar q_1 q_2 \bar q_3 q_4 | }{\bar q_5 }{b \bar q_6 q_7 | }{\bar q_8 }
\langle \bar q_1 q_2 \bar q_3 q_4 | \bar q_5 b \bar q_6 q_7 | \bar q_8 b \rangle
}~,
\nn\\
{\underline S}=
{
\contraction[1ex]{\langle \bar q_1 }{q_2 }{}{\bar q_3 }
\contraction[2ex]{\langle }{\bar q_1 }{ q_2 \bar q_3 }{q_4 }
\contraction[2ex]{\langle \bar q_1 q_2 \bar q_3 q_4 | }{\bar q_5 }{b \bar q_6 }{q_7 }
\contraction[1ex]{\langle \bar q_1 q_2 \bar q_3 q_4 | \bar q_5 b }{\bar q_6 }{q_7 | }{\bar q_8 }
\langle \bar q_1 q_2 \bar q_3 q_4 | \bar q_5 b \bar q_6 q_7 | \bar q_8 b \rangle
}~,
&~~
{\underline T}=
{
\contraction[1ex]{\langle }{\bar q_1 }{ q_2 \bar q_3 }{q_4 }
\contraction[1ex]{\langle \bar q_1 q_2 \bar q_3 q_4 | \bar q_5 b }{\bar q_6 }{q_7 | }{\bar q_8 }
\contraction[2ex]{\langle \bar q_1 }{q_2 }{\bar q_3 q_4 | }{\bar q_5 }
\contraction[3ex]{\langle \bar q_1 q_2 }{\bar q_3 }{q_4 | \bar q_5 b \bar q_6 }{q_7 }
\langle \bar q_1 q_2 \bar q_3 q_4 | \bar q_5 b \bar q_6 q_7 | \bar q_8 b \rangle
}~,
&~~
{\underline U}=
{
\contraction[1ex]{\langle }{\bar q_1 }{}{q_2 }
\contraction[1ex]{\langle \bar q_1 q_2 }{\bar q_3 }{}{q_4 }
\contraction[2ex]{\langle \bar q_1 q_2 \bar q_3 q_4 | }{\bar q_5 }{b \bar q_6 }{q_7 }
\contraction[1ex]{\langle \bar q_1 q_2 \bar q_3 q_4 | \bar q_5 b }{\bar q_6 }{q_7 | }{\bar q_8 }
\langle \bar q_1 q_2 \bar q_3 q_4 | \bar q_5 b \bar q_6 q_7 | \bar q_8 b \rangle
}~,
\nn\\
{\underline V}=
{
\contraction[1ex]{\langle \bar q_1 q_2 }{\bar q_3 }{}{q_4 }
\contraction[2ex]{\langle \bar q_1 }{q_2 }{\bar q_3 q_4 | }{\bar q_5 }
\contraction[3ex]{\langle }{\bar q_1 }{ q_2 \bar q_3 q_4 | \bar q_5 b \bar q_6 }{q_7 }
\contraction[1ex]{\langle \bar q_1 q_2 \bar q_3 q_4 | \bar q_5 b }{\bar q_6 }{q_7 | }{\bar q_8 }
\langle \bar q_1 q_2 \bar q_3 q_4 | \bar q_5 b \bar q_6 q_7 | \bar q_8 b \rangle
}~,
&~~
{\underline W}=
{
\contraction[1ex]{\langle \bar q_1 q_2 }{\bar q_3 }{}{q_4 }
\contraction[2ex]{\langle \bar q_1 }{q_2 }{\bar q_3 q_4 | }{\bar q_5 }
\contraction[1ex]{\langle \bar q_1 q_2 \bar q_3 q_4 | \bar q_5 b }{\bar q_6 }{}{q_7}
\contraction[3ex]{\langle }{\bar q_1 }{q_2 \bar q_3 q_4 | \bar q_5 b \bar q_6 q_7 | }{\bar q_8}
\langle \bar q_1 q_2 \bar q_3 q_4 | \bar q_5 b \bar q_6 q_7 | \bar q_8 b \rangle
}~,
&~~
{\underline X}=
{
\contraction[1ex]{\langle }{\bar q_1 }{ q_2 \bar q_3 }{q_4 }
\contraction[1ex]{\langle \bar q_1 q_2 \bar q_3 q_4 | \bar q_5 b }{\bar q_6 }{}{q_7}
\contraction[2ex]{\langle \bar q_1 }{q_2 }{\bar q_3 q_4 | }{\bar q_5 }
\contraction[3ex]{\langle \bar q_1 q_2 }{\bar q_3 }{q_4 | \bar q_5 b \bar q_6 q_7 | }{\bar q_8 }
\langle \bar q_1 q_2 \bar q_3 q_4 | \bar q_5 b \bar q_6 q_7 | \bar q_8 b \rangle
}~.
\label{eqcontract}
\eea

Not all the contractions are independent. For example, consider the
two contractions
\beq
{\underline E}(M_1M_2)=
{
\contraction[1ex]{\langle \bar q_1 q_2 \bar q_3 }{q_4 }{| \bar q_5 b }{\bar q_6 }
\contraction[2ex]{\langle \bar q_1 q_2 }{\bar q_3 }{q_4 | \bar q_5 b \bar q_6 }{q_7 }
\contraction[3ex]{\langle \bar q_1 }{q_2 }{\bar q_3 q_4 | }{\bar q_5 }
\contraction[4ex]{\langle }{\bar q_1 }{q_2 \bar q_3 q_4 | \bar q_5 b \bar q_6 q_7 | }{\bar q_8}
\langle \bar q_1 q_2 \bar q_3 q_4 | \bar q_5 b \bar q_6 q_7 | \bar q_8 b \rangle
}~,~~
{\underline M}( M_1 M_2)=
{
\contraction[1ex]{\langle \bar q_1 q_2 \bar q_3 }{q_4 }{| }{\bar q_5 }
\contraction[2ex]{\langle \bar q_1 }{q_2 }{\bar q_3 q_4 | \bar q_5 b }{\bar q_6 }
\contraction[3ex]{\langle }{\bar q_1 }{ q_2 \bar q_3 q_4 | \bar q_5 b \bar q_6 }{q_7 }
\contraction[4ex]{\langle \bar q_1 q_2 }{\bar q_3 }{q_4 | \bar q_5 b \bar q_6 q_7 | }{\bar q_8 }
\langle \bar q_1 q_2 \bar q_3 q_4 | \bar q_5 b \bar q_6 q_7 | \bar q_8 b \rangle
}~.
\label{tageqn}
\eeq
Here we label the contractions by $(M_1M_2)$ with the understanding
that $M_1 = \bar q_1 q_2$ and $M_2 =\bar q_3 q_4$. A contraction
labeled by $(M_2 M_1)$ would assume the assignment $M_2 = \bar q_1
q_2$ and $M_1 =\bar q_3 q_4$. Now, it is clear that the above
contractions are not independent as we have $ {\underline M}( M_1
M_2)={\underline E}(M_2 M_1)$.  This is just a consequence of the
final-state symmetry.

Applying the same final-state symmetry to all contractions, we get the
following equivalences between contractions:
\begin{center}
\begin{math}
{\hskip-0.5truein
\left(
\begin{array}{cccccccccccccccccccccccc}   
{\underline A} & {\underline B} & {\underline C} & {\underline D} & {\underline
E} & {\underline F} & {\underline G} & {\underline H} & {\underline I} &
{\underline J} & {\underline K} & {\underline L} & {\underline M} & {\underline
N} & {\underline O} & {\underline P} & {\underline Q} & {\underline R} &
{\underline S} & {\underline T} & {\underline U} & {\underline V} & {\underline
W} & {\underline X}\\
{\widetilde {\underline Q}} & { \wt {\underline R}} & { \wt {\underline O}}
&{\wt {\underline P}} & { \wt {\underline M}} & { \wt {\underline I}} & \wt
{{\underline X}} & { \wt {\underline T}} & { \wt {\underline F}} & {\wt
{\underline J}} & { \wt {\underline W}} &{ \wt {\underline V}} & { \wt
{\underline E}} & {\wt {\underline N}} & { \wt {\underline C}} & { \wt
{\underline D}} & { \wt {\underline A}} & {\wt {\underline B}} & {\wt
{\underline S}} & { \wt {\underline H}} & {\wt {\underline U}} & { \wt
{\underline L}} & { \wt {\underline K}} & { \wt {\underline G}}
\end{array}
\right)~.
}
\end{math}
\end{center}
Above, an element of the first line is specified by $(M_1 M_2)$ and is
equivalent to the corresponding element in the second line labeled by
$(M_2 M_1)$ and vice versa. Thus, any amplitude component can be be
written in terms of the contractions ${\underline A}$--${\underline X}$,
or equivalently in terms of the ``tilded'' contractions of the second
line. We therefore see that, of the 24 contractions, only 14 are
independent. These correspond to the 14 topologies of BS. We take the
14 independent contractions to be ${\underline A}$, ${\underline B}$,
${\underline C}$, ${\underline D}$, ${\underline E}$, ${\underline F}$,
${\underline G}$, ${\underline H}$, ${\underline J}$, ${\underline K}$,
${\underline L}$, ${\underline N}$, ${\underline S}$ and ${\underline U}$
(with or without tildes).

Note that care must be taken in using these equivalences. Above it is
assumed that the same operator is present in the two lines. If
different operators are involved, these equivalences only apply if
further symmetries are assumed (e.g.\ between $M_1$ and $M_2$).  This
will be important in Sec.~4.

It is useful to separate the 14 independent contractions into four
different classes. The advantage of this classification is that it
becomes easy to implement theory inputs and certain classes of
contractions can be argued to be small.

\begin{itemize}

\item {\bf Class I: emission topologies:} In these decays all the
quarks in $H_{eff}$, apart from the $b$ quark, are contracted with
quarks in the final state. The spectator quark has no contraction with
any quarks in $H_{eff}$. These contractions therefore have an
``inactive'' spectator quark. There are two such contractions,
${\underline E}$ and ${\underline F}$, which involve either an external or
internal emission of final-state mesons. We therefore rename these
with the `$EM$' label:
\beq
{\underline E} \to EM =
{
\contraction[1ex]{\langle \bar q_1 q_2 \bar q_3 }{q_4 }{| \bar q_5 b }{\bar q_6 }
\contraction[2ex]{\langle \bar q_1 q_2 }{\bar q_3 }{q_4 | \bar q_5 b \bar q_6 }{q_7 }
\contraction[3ex]{\langle \bar q_1 }{q_2 }{\bar q_3 q_4 | }{\bar q_5 }
\contraction[4ex]{\langle }{\bar q_1 }{q_2 \bar q_3 q_4 | \bar q_5 b \bar q_6 q_7 | }{\bar q_8}
\langle \bar q_1 q_2 \bar q_3 q_4 | \bar q_5 b \bar q_6 q_7 | \bar q_8 b \rangle
}~,~~
{\underline F} \to EM_{{\sss C}} =
{
\contraction[1ex]{\langle \bar q_1 }{q_2 }{\bar q_3 q_4 | }{\bar q_5 }
\contraction[2ex]{\langle \bar q_1 q_2 \bar q_3 }{q_4 }{| \bar q_5 b }{\bar q_6 }
\contraction[3ex]{\langle \bar q_1 q_2 }{\bar q_3 }{q_4 | \bar q_5 b \bar q_6 q_7 | }{\bar q_8 }
\contraction[4ex]{\langle }{\bar q_1 }{ q_2 \bar q_3 q_4 | \bar q_5 b \bar q_6 }{q_7 }
\langle \bar q_1 q_2 \bar q_3 q_4 | \bar q_5 b \bar q_6 q_7 | \bar q_8 b \rangle
} ~.
\eeq
The figures representing these contractions are given in
Fig.~\ref{figemission}.

\begin{figure}[ht]
\centerline{\epsfxsize=3.6truein \epsfbox{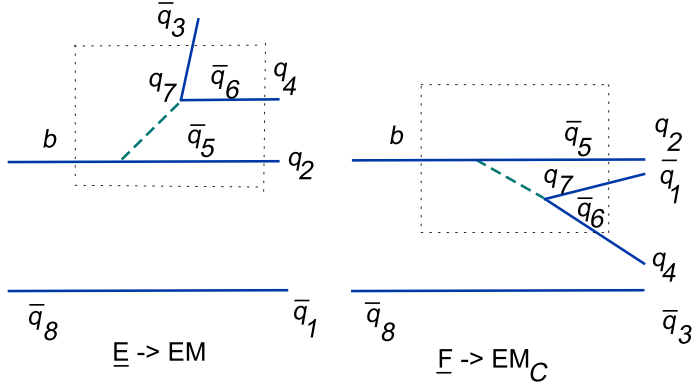}}
%\vskip-1truecm
\caption{Emission topologies. Though not shown, the quark lines are dressed with gluons.}
\label{figemission}
\end{figure}

Note that in factorization we have $EM \sim F_{\sss B \to M_1}f_{\sss
M_2}$, where $F_{\sss B \to M_1}$ is the $B \to M_1$ form factor and
$f_{\sss M_2}$ is the decay constant of $M_2$. Hence, in this
contraction $M_2$ is the emitted meson. Similarly, in factorization
$EM_{{\sss C}} \sim F_{\sss B \to M_2}f_{\sss M_1}$, and so here $M_1$
is the emitted meson. This contraction therefore corresponds to
colour-suppressed emission, and hence the subscript $C$ is used in the
label.

\item {\bf Class II: rescattering topologies:} In these contractions
the spectator quark does not contract with the quarks in $H_{eff}$ and
thus still remains ``inactive.'' However there is a contraction
between the quarks in $H_{eff}$. There are two possible contractions,
involving the pair $q_5,q_7$ (contractions ${\underline A}$ and
${\underline C}$) or the pair $q_6,q_7$ (contractions ${\underline G}$ and
${\underline K}$). For each pair the quarks in the final state can have
contractions between quark pairs belonging to different mesons
(${\underline A}$ and ${\underline G}$) or to the same meson (${\underline
C}$ and ${\underline K}$).  These latter contractions are expected to be
small as they are Okubo-Zweig-Iizuka (OZI) suppressed.

A possible ${\underline A}$-type contraction arises from the
rescattering contribution of the tree operators.  Since such
rescatterings are usually referred to as penguin contributions, we
rename this contraction $P$. Note that penguin operators can also
produce this type of contractions. We rename the ${\underline G}$
contraction as $PF$ since a Fierz transformation of the operator makes
the ${\underline G}$-type contraction look like an ${\underline A}$-type
contraction. Finally, the OZI-suppressed contractions are renamed
$P_{{\sss OZI}}$ and $PF_{{\sss OZI}}$:
\bea
{\underline A} \to P=
{
\contraction[1ex]{\langle \bar q_1 }{q_2 }{}{\bar q_3 }
\contraction[1ex]{\langle \bar q_1 q_2 \bar q_3 }{q_4 }{| \bar q_5 b }{\bar q_6 }
\contraction[2ex]{\langle \bar q_1 q_2 \bar q_3 q_4 | }{\bar q_5 }{b \bar q_6 }{q_7 }
\contraction[3ex]{\langle }{\bar q_1 }{q_2 \bar q_3 q_4 | \bar q_5 b \bar q_6 q_7 | }{\bar q_8}
\langle \bar q_1 q_2 \bar q_3 q_4 | \bar q_5 b \bar q_6 q_7 | \bar q_8 b \rangle
}~,
&
{\underline G} \to PF=
{
\contraction[1ex]{\langle \bar q_1 }{q_2 }{}{\bar q_3 }
\contraction[1ex]{\langle \bar q_1 q_2 \bar q_3 }{q_4 }{| }{\bar q_5 }
\contraction[1ex]{\langle \bar q_1 q_2 \bar q_3 q_4 | \bar q_5 b }{\bar q_6 }{}{q_7}
\contraction[2ex]{\langle }{\bar q_1 }{q_2 \bar q_3 q_4 | \bar q_5 b \bar q_6 q_7 | }{\bar q_8}
\langle \bar q_1 q_2 \bar q_3 q_4 | \bar q_5 b \bar q_6 q_7 | \bar q_8 b \rangle
}~,
\nn\\
{\underline C} \to P_{{\sss OZI}}=
{
\contraction[1ex]{\langle }{\bar q_1 }{}{q_2 }
\contraction[1ex]{\langle \bar q_1 q_2 \bar q_3 q_4 | }{\bar q_5 }{b \bar q_6 }{q_7 }
\contraction[2ex]{\langle \bar q_1 q_2 \bar q_3 }{q_4 }{| \bar q_5 b }{\bar q_6 }
\contraction[3ex]{\langle \bar q_1 q_2 }{\bar q_3 }{q_4 | \bar q_5 b \bar q_6 q_7 | }{\bar q_8 }
\langle \bar q_1 q_2 \bar q_3 q_4 | \bar q_5 b \bar q_6 q_7 | \bar q_8 b \rangle
}~,
&
{\underline K} \to PF_{{\sss OZI}}=
{
\contraction[1ex]{\langle }{\bar q_1 }{}{q_2 }
\contraction[1ex]{\langle \bar q_1 q_2 \bar q_3 }{q_4 }{| }{\bar q_5 }
\contraction[1ex]{\langle \bar q_1 q_2 \bar q_3 q_4 | \bar q_5 b }{\bar q_6 }{}{q_7}
\contraction[2ex]{\langle \bar q_1 q_2 }{\bar q_3 }{q_4 | \bar q_5 b \bar q_6 q_7 | }{\bar q_8 }
\langle \bar q_1 q_2 \bar q_3 q_4 | \bar q_5 b \bar q_6 q_7 | \bar q_8 b \rangle
}~.
\eea
The figures representing these contractions are given in
Fig.~\ref{figrescatt}.

\begin{figure}[ht]
\centerline{\epsfxsize=4.6truein \epsfbox{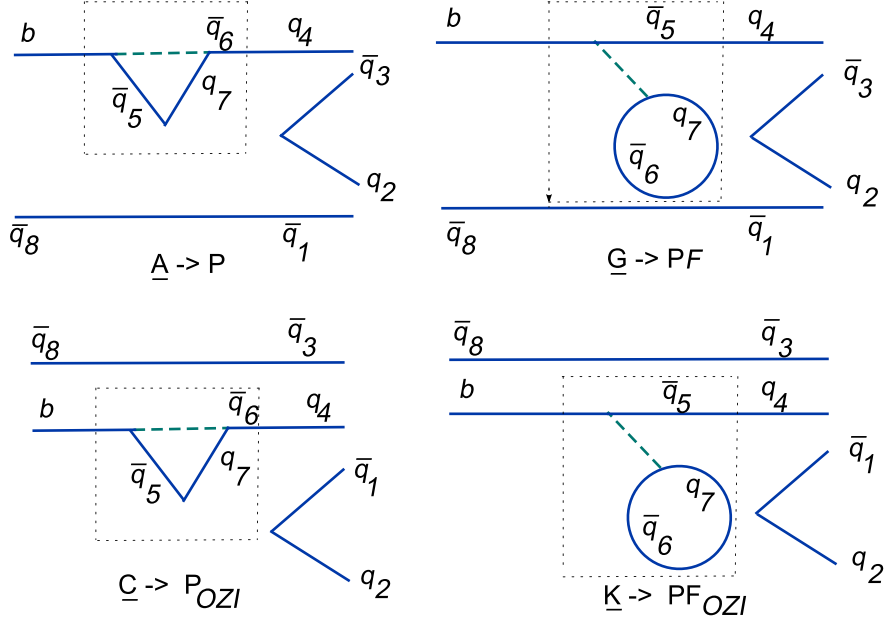}}
\caption{Rescattering topologies. Though not shown, the quark lines are dressed with gluons.}
\label{figrescatt}
\end{figure}

In the QCDfac and pQCD approaches these rescattering contributions are
perturbatively calculable. For a given operator such contributions are
suppressed by at least $\alpha_s(m_b) $ relative to the contractions
of the same operator belonging to Class I. For contractions involving
charm quarks it is possible that rescattering may involve
long-distance contributions that are not calculable perturbatively
\cite{SCET,charming}. To get an estimate of the size of the
rescattering contributions from charm intermediate states through $b
\to c \bar{c} s$ tree operators we can study charmless $B$
decays. Here there are no emission contractions for these operators;
they can contribute only through rescattering contractions.  Now, from
experiments we know that these rescattering contractions are small and
not of $O(1)$. If this were not the case, this would lead to too-large
branching fractions for charmless $b \to s$ transitions such as $B \to
K \pi$, $B \to \phi \ks$ etc. The rescattering from $b \to c \bar{c}
s$ tree operators is then typically of the size of penguin amplitudes
in charmless $b \to s$ transitions. Using flavour $SU(3)$ symmetry,
one can argue that rescattering through $b \to c \bar{c} d$ is also
small.

Hence, for a given decay and a given operator, the emission
contractions of class I, which do not suffer any colour suppression,
are generally larger than the rescattering contractions generated by
the same operator. To take a specific final state as an example, we
expect the colour-allowed decay $B \to D_{s}D$ to get its dominant
contribution from emission topologies.  Such arguments can also be
applied to NP operators and have been used to argue for small NP
strong phases from rescattering contractions \cite{Datta-Lon}.

As noted above, the OZI-suppressed contractions are expected to be
small. In fact, the $P_{{\sss OZI}}$-type contraction can contribute to
the decay $\bd \to J/\psi \ks$ with a weak phase different from that
of the dominant contribution through $u$-quark rescattering. If such
contributions were significant, the result would affect the $\sin
2\beta$ measurement using this mode. The fact that the measurement
\cite{sin2beta} does appear to agree with predictions strengthens the
claim that the OZI rescattering contributions are small.

\item{\bf Class III: annihilation/exchange topologies:} We identify
the annihilation and exchange topologies with the $({\underline
B},{\underline D})$ and $({\underline H},{\underline L})$ contractions,
respectively. ${\underline D}$ and ${\underline L}$ involve contractions
between quarks in the same mesons, and are therefore
OZI-suppressed. We therefore rename the annihilation contractions as
$A$ and $A_{\sss OZI}$, and the exchange contractions as $EX$ and
$EX_{{\sss OZI}}$:
\bea
{\underline B} \to A=
{
\contraction[1ex]{\langle \bar q_1 }{q_2 }{}{\bar q_3 }
\contraction[1ex]{\langle \bar q_1 q_2 \bar q_3 }{q_4 }{| \bar q_5 b }{\bar q_6 }
\contraction[2ex]{\langle }{\bar q_1 }{ q_2 \bar q_3 q_4 | \bar q_5 b \bar q_6 }{q_7 }
\contraction[3ex]{\langle \bar q_1 q_2 \bar q_3 q_4 | }{\bar q_5 }{b \bar q_6 q_7 | }{\bar q_8 }
\langle \bar q_1 q_2 \bar q_3 q_4 | \bar q_5 b \bar q_6 q_7 | \bar q_8 b \rangle
}~,
&
{\underline H} \to EX=
{
\contraction[1ex]{\langle \bar q_1 }{q_2 }{}{\bar q_3 }
\contraction[1ex]{\langle \bar q_1 q_2 \bar q_3 }{q_4 }{| }{\bar q_5 }
\contraction[1ex]{\langle \bar q_1 q_2 \bar q_3 q_4 | \bar q_5 b }{\bar q_6 }{q_7 | }{\bar q_8 }
\contraction[2ex]{\langle }{\bar q_1 }{ q_2 \bar q_3 q_4 | \bar q_5 b \bar q_6 }{q_7 }
\langle \bar q_1 q_2 \bar q_3 q_4 | \bar q_5 b \bar q_6 q_7 | \bar q_8 b \rangle
}~,
\nn\\
{\underline D} \to A_{{\sss OZI}}=
{
\contraction[1ex]{\langle }{\bar q_1 }{}{q_2 }
\contraction[1ex]{\langle \bar q_1 q_2 \bar q_3 }{q_4 }{| \bar q_5 b }{\bar q_6 }
\contraction[2ex]{\langle \bar q_1 q_2 \bar q_3 q_4 | }{\bar q_5 }{b \bar q_6 q_7 | }{\bar q_8 }
\contraction[3ex]{\langle \bar q_1 q_2 }{\bar q_3 }{q_4 | \bar q_5 b \bar q_6 }{q_7 }
\langle \bar q_1 q_2 \bar q_3 q_4 | \bar q_5 b \bar q_6 q_7 | \bar q_8 b \rangle
}~,
&
{\underline L} \to EX_{{\sss OZI}}
{
\contraction[1ex]{\langle }{\bar q_1 }{}{q_2 }
\contraction[1ex]{\langle \bar q_1 q_2 \bar q_3 }{q_4 }{| }{\bar q_5 }
\contraction[1ex]{\langle \bar q_1 q_2 \bar q_3 q_4 | \bar q_5 b }{\bar q_6 }{q_7 | }{\bar q_8 }
\contraction[2ex]{\langle \bar q_1 q_2 }{\bar q_3 }{q_4 | \bar q_5 b \bar q_6 }{q_7 }
\langle \bar q_1 q_2 \bar q_3 q_4 | \bar q_5 b \bar q_6 q_7 | \bar q_8 b \rangle
}~.
\eea
These are shown in Fig.~\ref{figannex}.

\begin{figure}[ht]
\centerline{\epsfxsize=4.2truein \epsfbox{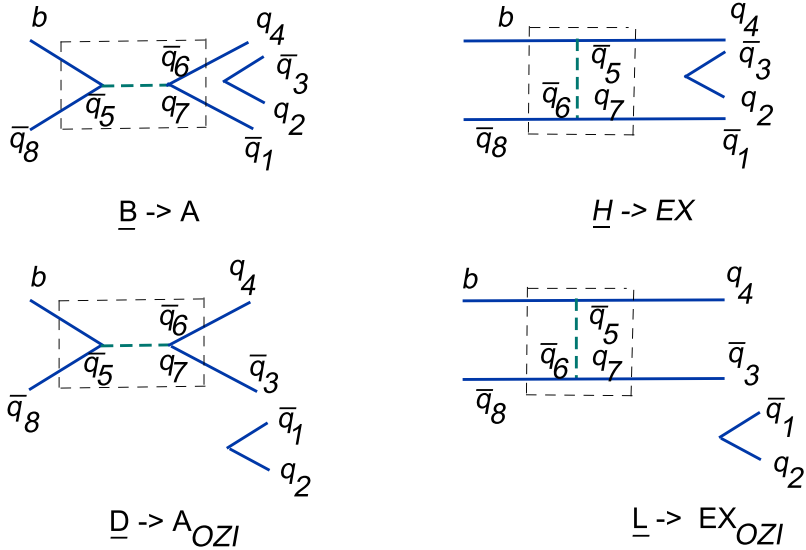}}
\caption{Annihilation/exchange topologies. Though not shown, the quark lines are dressed with gluons.}
\label{figannex}
\end{figure}

This class of contractions is suppressed by $O(1/m_b)$. Whether such
contributions are sufficiently small is a matter of debate which will
ultimately be settled by observing (or not) decays such as $\bd \to
K^+ K^-$ or $\bd \to D_s^+ D_s^-$, which can only proceed through
annihilation and exchange diagrams.

The OZI-suppressed contractions in this class are again expected to be
small. Their size can be obtained from the measurement of decays such
as $\bd \to J/\psi \phi$ \cite{dattalipkin} which can go through such
OZI-suppressed contractions.

\item{\bf Class IV: topologies with annihilation/exchange $+$
rescattering:} This class includes contractions that have annihilation
and exchange in combination with rescattering. These include
${\underline N}$, ${\underline S}$, ${\underline J}$ and ${\underline U}$. The
${\underline N}$ contraction can be renamed $PA$ for it is a
``penguin-annihilation'' contraction. Similarly, ${\underline S}$ is a
``penguin-exchange'' contraction: $PE$. Note that $PA$ and $PE$ are
OZI-suppressed while the contractions ${\underline U}$ and ${\underline
J}$ suffer an additional OZI suppression; we rename them as $PA_{{\sss
OZI}}$ and $PE_{{\sss OZI}}$ (Fig.~\ref{figannexrescatt}):
\bea
{\underline N} \to PA=
{
\contraction[1ex]{\langle \bar q_1 }{q_2 }{}{\bar q_3 }
\contraction[2ex]{\langle }{\bar q_1 }{ q_2 \bar q_3 }{q_4 }
\contraction[1ex]{\langle \bar q_1 q_2 \bar q_3 q_4 | \bar q_5 b }{\bar q_6 }{}{q_7}
\contraction[2ex]{\langle \bar q_1 q_2 \bar q_3 q_4 | }{\bar q_5 }{b \bar q_6 q_7 | }{\bar q_8 }
\langle \bar q_1 q_2 \bar q_3 q_4 | \bar q_5 b \bar q_6 q_7 | \bar q_8 b \rangle
}~,
&
{\underline S} \to PE=
{
\contraction[1ex]{\langle \bar q_1 }{q_2 }{}{\bar q_3 }
\contraction[2ex]{\langle }{\bar q_1 }{ q_2 \bar q_3 }{q_4 }
\contraction[2ex]{\langle \bar q_1 q_2 \bar q_3 q_4 | }{\bar q_5 }{b \bar q_6 }{q_7 }
\contraction[1ex]{\langle \bar q_1 q_2 \bar q_3 q_4 | \bar q_5 b }{\bar q_6 }{q_7 | }{\bar q_8 }
\langle \bar q_1 q_2 \bar q_3 q_4 | \bar q_5 b \bar q_6 q_7 | \bar q_8 b \rangle
}~,
\nn\\
{\underline J} \to PA_{{\sss OZI}}
{
\contraction[1ex]{\langle }{\bar q_1 }{}{q_2 }
\contraction[1ex]{\langle \bar q_1 q_2 }{\bar q_3 }{}{q_4 }
\contraction[2ex]{\langle \bar q_1 q_2 \bar q_3 q_4 | }{\bar q_5 }{b \bar q_6 q_7 | }{\bar q_8 }
\contraction[1ex]{\langle \bar q_1 q_2 \bar q_3 q_4 | \bar q_5 b }{\bar q_6 }{}{q_7}
\langle \bar q_1 q_2 \bar q_3 q_4 | \bar q_5 b \bar q_6 q_7 | \bar q_8 b \rangle
}~,
&
{\underline U} \to PE_{{\sss OZI}}=
{
\contraction[1ex]{\langle }{\bar q_1 }{}{q_2 }
\contraction[1ex]{\langle \bar q_1 q_2 }{\bar q_3 }{}{q_4 }
\contraction[2ex]{\langle \bar q_1 q_2 \bar q_3 q_4 | }{\bar q_5 }{b \bar q_6 }{q_7 }
\contraction[1ex]{\langle \bar q_1 q_2 \bar q_3 q_4 | \bar q_5 b }{\bar q_6 }{q_7 | }{\bar q_8 }
\langle \bar q_1 q_2 \bar q_3 q_4 | \bar q_5 b \bar q_6 q_7 | \bar q_8 b \rangle
}~,
\
\label{rescatt-ann-exch}
\eea

\begin{figure}[ht]
\centerline{\epsfxsize=3.6truein \epsfbox{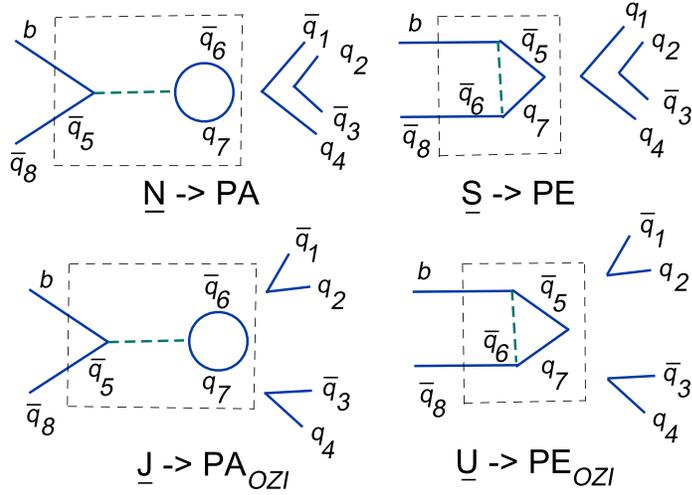}}
%\vskip-1truecm
\caption{Topologies with annihilation/exchange $+$
rescattering. Though not shown, the quark lines are dressed with
gluons.}
\label{figannexrescatt}
\end{figure}

The contractions in this class are expected to be tiny as they are
both $O(1/m_b)$-suppressed and OZI suppressed.

\end{itemize}

To summarize: there are 14 independent contractions. These can be
separated into four classes: $\left\{EM, ~EM_{{\sss C}}\right\}$,
$\left\{P, ~PF, ~P_{{\sss OZI}}, ~PF_{{\sss OZI}}\right\}$, $\left\{A,
~EX, ~A_{{\sss OZI}}, ~EX_{{\sss OZI}}\right\}$, \\ $\left\{PA, ~PE,
~PA_{{\sss OZI}}, ~PE_{{\sss OZI}}\right\}$. In what follows, we will
almost always express amplitudes in terms of these contractions.
However, we will occasionally write amplitudes in terms of the 24
contractions ${\underline A}$--${\underline X}$ [Eq.~(\ref{eqcontract})].
When we do this, we will be sure to give a warning of this fact. Thus,
unless there is an explicit statement to the contrary, the reader can
assume that all amplitudes are given in terms of the contractions
$EM$, $EM_{{\sss C}}$, etc.

Buras-Silvestrini (BS) also have 14 topologies, described by rather
complicated, nonstandard figures. They label these as DE (Disconnected
Emission), CE (Connected Emission), DA (Disconnected Annihilation), CA
(Connected Annihilation), DEA (Disconnected Emission-Annihilation),
CEA (Connected Emission-Annihilation), DP (Disconnected Penguin), CP
(Connected Penguin), DPE (Disconnected Penguin-Emission), CPE
(Connected Penguin-Emission), DPA (Disconnected Penguin-Annihilation),
CPA (Connected Penguin-Annihilation), ${\overline{\rm DPA}}$
(Disconnected Double-Penguin-Annihilation), ${\overline{\rm CPA}}$
(Connected Double-Penguin-Annihilation). Their eight OZI-suppressed
topologies are DEA, CEA, DPE, CPE, DPA, CPA, ${\overline{\rm DPA}}$,
${\overline{\rm CPA}}$. As such, the starting point of our method is
very similar to theirs. However, what they do is rather different from
what we do in this paper.

BS's main aim is to produce a manifestly scheme- and scale-independent
formalism for hadronic $B$ decays. In the subsequent sections we show
that this method reproduces results which have been obtained using
other techniques. In particular, the results of diagrams, GPY, etc.\
can be simply expressed in terms of contractions.

BS stress the need to include the ``small'' OZI-suppressed
contractions in order to obtain scheme- and scale-independent
results. We agree with this observation, in principle. However, the BS
paper was written before the observation that certain contractions are
suppressed by factors of $O(1/m_b)$.  Thus, in practice, these
contributions are indeed small, and it is reasonable to neglect them
when computing the amplitudes for hadronic $B$ decays. Including the
small contractions makes the expressions for the amplitudes
unnecessarily complicated.

Finally, BS show how to write diagrams in terms of their contractions
(Sec.~10 of their paper). However, these diagrams are not those
described in Ref.~\cite{GHLR} and used elsewhere. In the present paper
we show how the diagrams of Ref.~\cite{GHLR} can be written simply in
terms of contractions.

\subsection{$\btopik$ Decays}

In this subsection, we show how to compute the possible contractions
for a given $B\to M_1 M_2$ decay. We do so by considering the four
$\btopik$ decays: $B^- \to \pi^- \bar K^0$, $B^- \to \pi^0 K^-$, $\bar
B^0 \to \pi^+ K^-$, $\bar B^0 \to \pi^0 \bar K^0$.

We begin with $B^- \to \pi^- \bar K^0$. The operator
$\mathcal{H}_{\sss T}$ that generates tree amplitudes involves $O_1$
and $O_2$. Then
\bea
T^{-0} &=& \frac{1}{\sqrt{2}} \sum_{contractions} \left( \bra{\pi^-
\bar K^0} \mathcal{H}_{\sss T} \ket{B^-} + \bra{\bar K^0 \pi^-} \mathcal{H}_{\sss T}
\ket{B^-} \right)\nn\\
&=& \lambda_u^{(s)} c_1(T^{-0})^u_1 + \lambda_c^{(s)}c_1(T^{-0})^c_1 +
\lambda_u^{(s)}c_2(T^{-0})^u_2 + \lambda_c^{(s)}c_2(T^{-0})^c_2~,
\label{eq1}
\eea 
where
\beq
(T^{-0})^p_i = \frac{1}{\sqrt{2}} \sum_{contractions} \left(
\bra{\pi^- \bar K^0} O^p_i \ket{B^-} + \bra{\bar K^0 \pi^-} O^p_i
\ket{B^-} \right)~.
\eeq
For $O^u_1$:
\bea
\label{conAQ}
(T^{-0})^u_1 & = & \frac{1}{\sqrt{2}} \sum_{contractions} \left(
\bra{\bar u d \bar d s} \bar u b\bar s u \ket{\bar u b} + \bra{\bar d
s \bar u d} \bar u b\bar s u \ket{\bar u b} \right) \nn\\
& = & \frac{1}{\sqrt{2}} \left[ ({\underline A}^{\prime u}_1(\pi^- \bar K^0) +
{\underline B}^{\prime u}_1(\pi^- \bar K^0)) + ({\wt {\underline Q}}^{\prime u}_1( \bar K^0
\pi^-) + {\wt{\underline R}}^{\prime u}_1( \bar K^0 \pi^-) )\right] ~.
\eea
In the above, we have expressed in the amplitude in terms of the
contractions ${\underline A}$-${\underline X}$ of
Eq.~(\ref{eqcontract}). Here and below, the primes on the contractions
indicate a $\btos$ transition.

Now, final-state symmetry indicates that ${\tilde {\underline
Q}}^{\prime u}_1( \bar K^0 \pi^-) = {\underline A}^{\prime u}_1(\pi^-
\bar K^0)$ and ${\tilde {\underline R}}^{\prime u}_1( \bar K^0 \pi^-) =
{\underline B}^{\prime u}_1(\pi^- \bar K^0)$. Using this, and changing
the notation to that used in the four contraction classes, we then
obtain for $B^- \to \pi^- \bar K^0$ (dropping the $(M_1 M_2)$ label):
\beq 
(T^{-0})^u_1 = \sqrt{2} (P^{\prime u}_1 + A^{\prime u}_1)~.  
\eeq
(Here, $P'$ and $A'$ indicate penguin and annihilation contractions,
respectively.) In the above we have shown how final-state symmetry is
used to include the ${\tilde {\underline Q}}$ and ${\tilde {\underline
R}}$ contractions. From here on, final-state symmetry will be assumed
and not shown explicitly.

For $O^c_1$:
\beq
(T^{-0})^c_1 = \frac{1}{\sqrt{2}} \sum_{contractions} \left( \bra{\bar
u d \bar d s} \bar c b\bar s c \ket{\bar u b} + \bra{\bar d s \bar u
d} \bar c b\bar s c \ket{\bar u b} \right)~.
\eeq
Only the $P'$ contraction is allowed and we get
\beq
(T^{-0})^c_1 = \sqrt{2} P^{\prime c}_1~.
\eeq

For $O^u_2$ and $O^c_2$, only the colour structure is different, so that
\bea
(T^{-0})^u_2 &=& \sqrt{2} (P^{\prime u}_2 + A^{\prime u}_2)~,\nn\\
(T^{-0})^c_2 &=& \sqrt{2} P^{\prime c}_2~,
\eea
giving
\beq 
T^{-0} = \sqrt{2} c_i [\lambda_u^{(s)} (P^{\prime u}_i + A^{\prime
u}_i) + \lambda_c^{(s)} P^{\prime c}_i]~,
\eeq
where a sum over $i=1,2$ is understood. 

The tree pieces of the other three $\btopik$ decays can be found in
the same way. In all, we have 
\bea
T^{-0} &=& \sqrt{2}c_i [\lambda_u^{(s)} (P^{\prime u}_i + A^{\prime
u}_i) + \lambda_c^{(s)} P^{\prime c}_i]~,\nn\\
T^{0-} &=& c_i [-\lambda_u^{(s)} (P^{\prime u}_i + A^{\prime u}_i +
EM^{\prime u}_i +EM^{\prime u}_{{\sss C} i}) - \lambda_c^{(s)} P^{\prime
c}_i]~,\nn\\
T^{+-} &=& \sqrt{2}c_i [-\lambda_u^{(s)} (P^{\prime u}_i+EM^{\prime
u}_i) - \lambda_c^{(s)} P^{\prime c}_i]~,\nn\\
T^{00} &=& c_i [\lambda_u^{(s)} (P^{\prime u}_i -EM^{\prime u}_{{\sss C} i}) +
\lambda_c^{(s)} P^{\prime c}_i]~.
\label{ampltrees}
\eea
Note that the quadrilateral isospin relation for $\btopik$ decays is
respected:
\beq
T^{-0} + \sqrt{2} T^{0-} = T^{+-} + \sqrt{2}T^{00}~.
\eeq

We now turn to the EWP operators. We make the approximation of
neglecting $C_7$ and $C_8$, so that the EWP operators are purely
$(V-A)\times(V-A)$ and are given by $O_9$ and $O_{10}$ alone. 

In these operators, one sums over the light quarks, multiplied by
their charge. Below we consider $q'=u,d,s$ (with $e_u = 2/3$, $e_d =
e_s = -1/3$).  Below we will always assume isospin symmetry, so that
the contractions of $u$ and $d$ quarks are equal. However, we do not
assume flavour SU(3) symmetry here (under SU(3) the contractions of
$u$, $d$ and $s$ quarks are equal).  Instead the $u$/$d$ and $s$
contractions are labeled individually.

The EWP contractions for the four $\btopik$ decays can be found in the
same way as was done for the tree operators. We have
\bea
P_{\sss EW}^{-0} &=& \sqrt{2} \frac{3}{2} \lambda_t^{(s)} c_i \left[
 -\frac{1}{3} { PF}_i^\prime - \frac{2}{3} { EX}_i^\prime + \frac{1}{3}
 {\wt{EM}}_{{\sss C} i}^\prime + \frac13 \left( { P}_i^{\prime s} + { PF}_i^{\prime s} \right)
 \right]~,\nn\\
P_{\sss EW}^{0-} &=& \frac{3}{2} \lambda_t^{(s)} c_i \left[
\frac{1}{3} { PF}_i^\prime + { {\widetilde {EM}}}_i^\prime +\frac{2}{3} { EX}_i^\prime +
\frac{2}{3} {\wt{EM}}_{{\sss C} i}^\prime - \frac13 \left( { P}_i^{\prime s} 
+ { PF}_i^{\prime
s} \right) \right]~,\nn\\
P_{\sss EW}^{+-} &=& \sqrt{2} \frac{3}{2} \lambda_t^{(s)} c_i \left[
\frac{1}{3} { PF}_i^\prime -\frac{1}{3} { EX}_i^\prime + \frac{2}{3}
{\wt{EM}}_{{\sss C} i}^\prime - \frac13 \left( { P}_i^{\prime s} + { PF}_i^{\prime s} \right)
\right]~,\nn\\
P_{\sss EW}^{00} &=& \frac{3}{2} \lambda_t^{(s)} c_i \left[
 -\frac{1}{3} { PF}_i^\prime + {\wt{EM}}_i^\prime +\frac{1}{3} {\bar EX}_i^\prime +
 \frac{1}{3} {\wt{EM}}_{{\sss C} i}^\prime + \frac13 \left( { P}_i^{\prime s} 
 + { PF}_i^{\prime
  s} \right) \right]~.
\label{amplewp}
\eea
Here, a sum over $i=9,10$ is understood. Note the appearance of the
terms ${\widetilde {EM}}_i^\prime$ and ${\wt{EM}}_{{\sss C}
i}^\prime$. We remind the reader that the tilde indicates that the
final state is $M_2 M_1$ as opposed to $M_1 M_2$. Thus, for $\btopik$
decays, the EWP contractions involve final states of the form $K\pi$.
(To be more precise, EWP contractions involve $\pi K$ terms which are
equivalent to $K\pi$ contractions, as per the equivalence table
following Eq.~(\ref{tageqn}).)

In the above, $u$-quark and $d$-quark contractions are taken to be
equal (isospin symmetry). The $u$/$d$ contractions have no label,
while the contractions of $s$-quarks are labeled by an `$s$'
superscript.  Because the operators in the two sets of contractions
are different, the quarks are in different positions, and the
equivalences due to final-state symmetry do not apply here. On the
other hand, in the SU(3) limit, $u$-quark, $d$-quark and $s$-quark
contractions are equal, and the effect of SU(3) symmetry will be
explicitly worked out in Sec.~4. One sees that, once again, the
isospin quadrilateral is respected above.

We note in passing that there are rescattering contractions involving
the $c$ quark for the EWP penguin operators which are represented by
$P_{9,10}^{'c}$ and $PF_{9,10}^{'c}$. However these contributions are
expected to be tiny relative to the rescattering from the tree
operators. This is unlike the case of rescattering contractions
involving the $u$ quark where the EWP contractions are enhanced by CKM
factors relative to the tree contractions, so that both contractions
are of the same size.

Finally, we turn to the gluonic-penguin operators. Here no WC's can be
neglected, so that all operators $O_3$-$O_6$ must be included (i.e.\
we take into account both $(V-A)\times(V-A)$ and $(V-A)\times(V+A)$
Dirac structures). We have
\bea
P^{-0} &=& - \sqrt{2}\lambda_t^{(s)} c_i \left[ 2 {PF}_i^{\prime}
+ {PF}_i^{\prime s} + {EX}_i^{\prime} + {\wt{EM}}_{{\sss C} i}^{\prime} +
{P}_i^{\prime s} \right] ~,\nn\\
P^{0-} &=& \sqrt{2}\lambda_t^{(s)} c_i \left[ 2 {PF}_i^{\prime} +
{PF}_i^{\prime s} + {EX}_i^{\prime} + {\wt{EM}}_{{\sss C} i}^{\prime} +
{P}_i^{\prime s} \right] ~,\nn\\
P^{+-} &=& \sqrt{2}\lambda_t^{(s)} c_i \left[ 2 {PF}_i^{\prime} +
{PF}_i^{\prime s} + {EX}_i^{\prime} + {\wt{EM}}_{{\sss C} i}^{\prime} +
{\bar P}_i^{\prime s} \right] ~,\nn\\
P^{00} &=& - \sqrt{2} \lambda_t^{(s)}\sqrt{2} c_i \left[ 2 { PF}_i^{\prime}
+ {PF}_i^{\prime s} + {EX}_i^{\prime} + {\wt{EM}}_{{\sss C} i}^{\prime} +
{\bar P}_i^{\prime s} \right] ~,
\label{amplping}
\eea
which respects the isospin quadrilateral relation. The sum is over
$i=3,4,5,6$, and the index `$i$' indicates the Dirac structure.

In the above, we have made no assumptions about colours. That is, the
quarks can have any colour, representing the exchange of any number of
gluons. Thus, the above equations hold to all orders in $\alpha_s$.
(We will work order-by-order in Sec.~5.)

In this subsection, we have shown how to write the matrix elements for
the four $\btopik$ decays in terms of contractions. This can be done
for any hadronic $B$ decay. (Indeed, this is essentially what BS have
done.) One important point is that, while we have expressed matrix
elements in terms of contractions, we have not evaluated these matrix
elements. This requires an additional method, such as QCDfac or pQCD.

\section{Connection to Diagrams}

All $\btopik$ decays receive contributions from the penguin diagram,
$P'$. (As with contractions, a prime on a diagram indicates a $\btos$
transition.) This diagram actually contains three pieces,
corresponding to the identity of the internal quark:
\beq
P' = \lambda^{(s)}_u P'_u + \lambda^{(s)}_c P'_c + \lambda^{(s)}_t
P'_t ~.
\eeq
Below we use this form for $P'$. Note that the unitarity of the
Cabibbo-Kobayashi-Maskawa (CKM) matrix has {\it not} been used. This
is because the unitarity of the CKM matrix was not used in writing the
effective Hamiltonian [Eq.~(\ref{Heff})].

When writing the $\btopik$ amplitudes in terms of diagrams, it is
conventional to absorb all $\lambda^{(s)}_p$ ($p=u,c,t$) factors into
the diagrams themselves. We then have
\bea
\label{ampsdiags}
A^{-0} & = & [P'_u + A'] + [P'_c] + [ P'_t - \frac13 \pewcp +
\frac{2}{3} \pewep + \frac{1}{3} \pewpup] ~, \nn\\
\sqrt{2} A^{0-} & = & [ -T' -C' - P'_u - A' ] + [-P'_c] \nn\\ & &
\hskip1truecm +~ [ -P'_t -\pewp -\frac23 \pewcp -\frac{2}{3} \pewep
-\frac{1}{3}\pewpup] ~, \\
A^{+-} & = & [ -T' - P'_u] + [-P'_c] + [ -P'_t - \frac23 \pewcp
+\frac{1}{3} \pewep -\frac{1}{3} \pewpup] ~, \nn\\
\sqrt{2}A^{00} & = & [ -C' +P'_u ] + [P'_c] + [ P'_t -\pewp -
  \frac13 \pewcp -\frac{1}{3} \pewep + \frac{1}{3} \pewpup] ~. \nn
\eea
In Ref.~\cite{GPY} it was shown that there is an EWP diagram
corresponding to each of the $T$, $C$, $P_u$, $A$, $E$, $PA_u$
diagrams. These are included above. The first, second and third terms
in square brackets are proportional to $\lambda^{(s)}_u$,
$\lambda^{(s)}_c$ and $\lambda^{(s)}_t$, respectively.

In the previous section it was shown how to write the contributions
from tree, EWP and gluonic-penguin operators in terms of
contractions. Using the above expressions, we can now write diagrams
in terms of contractions. The tree-operator contribution is given in
Eq.~(\ref{ampltrees}). Comparing with the first and second terms in
parentheses above, we obtain
\bea
P'_u &=& \sqrt{2} \lambda_u^{(s)} c_i P^{\prime u}_i~,\nn\\
P'_c &=& \sqrt{2} \lambda_c^{(s)} c_i P^{\prime c}_i~,\nn\\ 
T' &=& \sqrt{2} \lambda_u^{(s)} c_i EM^{\prime u}_i~,\nn\\
C' &=& \sqrt{2} \lambda_u^{(s)} c_i EM^{\prime u}_{{\sss C} i}~,\nn\\
A' &=& \sqrt{2} \lambda_u^{(s)} c_i A^{\prime u}_i~,
\label{treediagcont}\eea
where the sum over $i=1,2$ is still understood. This matching looks
very natural. Graphically, an $EM'$ emission contraction really looks
like a $T'$ diagram, and similarly for the colour-suppressed emission
contraction, $EM'_{{\sss C}}$, and $C'$ (Fig.~\ref{figemission}).
$P'_u$ and $P'_c$ are both described by the same type of contraction
(Fig.~\ref{figrescatt}), which represents rescattering from the
operators $O_{1,2}$. Finally, the $A'$ contraction can be identified
with annihilation diagram (Fig.~\ref{figannex}).

The contribution from EWP operators is given in Eq.~(\ref{amplewp}).
Comparing to the EWP terms above, we have
\bea
\pewp &=& -\sqrt{2} \, \frac{3}{2} \lambda_t^{(s)} c_i {\wt {EM}}'_{i}~,\nn\\
\pewcp &=& -\sqrt{2} \, \frac{3}{2} \lambda_t^{(s)} c_i {\wt {EM}}'_{{\sss C} i}~,\nn\\
\pewpup &=& -\sqrt{2} \, \frac{3}{2} \lambda_t^{(s)} c_i \left[
\left( {PF}^{\prime s}_i - {PF}'_i \right) + {P}^{\prime s}_i \right] ~,\nn\\
\pewep &=& -\sqrt{2} \, \frac{3}{2} \lambda_t^{(s)} c_i {EX}'_i~.
\label{ewpdiagcont}
\eea

Finally, the gluonic-penguin contribution is given in
Eq.~(\ref{amplping}). This implies
\beq
P'_t = -\sqrt{2} \lambda_t^{(s)} c_i \left[ 2 {PF}_i^{\prime} +
{PF}_i^{\prime s} + {EX}_i^{\prime} + {\wt {EM}}_{{\sss C} i}^{\prime} +
{\bar P}_i^{\prime s} \right] ~.
\label{pingdiagcont}
\eeq

Now, we note that Eqs.~(\ref{treediagcont}), (\ref{ewpdiagcont}) and
(\ref{pingdiagcont}) above are missing some diagrams: $E$,
$PA_{u,c,t}$, $\pewa$ and $\pewpau$. In order to obtain these, we must
use a different decay. Here we choose the three $\btopipi$ decays:
$B^- \to \pi^- \pi^0$, $\bar B^0 \to \pi^+ \pi^-$, $\bar B^0 \to \pi^0
\pi^0$ (recall that isospin symmetry has been assumed, so that the
placement of the final-state pions is unimportant).

We simply present the results here; they can be derived using the
techniques described previously. The tree contractions are
\bea
T^{\pi^-\pi^0} &=& -\lambda_u^{(d)} c_i (EM_i^u + EM_{{\sss C} i}^u) ~,\nn\\
T^{\pi^+\pi^-} &=& -\sqrt{2} c_i[\lambda_u^{(d)}  ( EM_i^u +P_i^u +
EX_i^u + PE_i^u) + \lambda_c^{(d)} (P_i^c + PE_i^c) ]~,\nn\\
T^{\pi^0\pi^0} &=&  c_i[\lambda_u^{(d)} ( - EM_{{\sss C} i}^u + P_i^u+ EX_i^u +
PE_i^u) + \lambda_c^{(d)} (P_i^c + PE_i^c) ]~.
\label{ampltreespipi}
\eea
The EWP contractions are
\bea
P_{\sss EW}^{\pi^-\pi^0} &=& \frac{3}{2} \lambda_t^{(d)} c_i ({ EM}_i + { EM}_{{\sss C} i}) ~,\nn\\
P_{\sss EW}^{\pi^+\pi^-} &=& \frac{3}{2} \sqrt{2} \lambda_t^{(d)} c_i 
\left[ \frac23 { EM}_{{\sss C} i} + \frac13 \left( { A}_i + {PF}_i + {PA}_i \right) \right. \nn\\
& & \hskip3truecm \left.
-~\frac13
\left( { P}_i + {EX}_i + {PE}_i \right) - \frac13 \left( {PF}_i^s + {PA}_i^s \right)
\right] ~, \nn\\
P_{\sss EW}^{\pi^0\pi^0} &=& \frac{3}{2} \lambda_t^{(d)} c_i \left[
{ EM}_i + \frac13 { EM}_{{\sss C} i} - \frac13 \left( {A}_i + { PF}_i + {PA}_i \right)
\right. \nn\\
& & \hskip3truecm \left. +~\frac13 \left( {P}_i + {EX}_i + {PE}_i \right) +
\frac13 \left( {PF}_i^s + {PA}_i^s \right) \right] ~.
\label{amplewppipi}
\eea
The contractions from the gluonic-penguin operators are
\bea
P^{\pi^-\pi^0} &=& 0~,\nn\\
P^{\pi^+\pi^-} &=& \sqrt{2} \lambda_t^{(d)} c_i \left[ \left( 2 {
PF}_i + {PF}_i^s + {EX}_i + {\wt EM}_{{\sss C} i} + { P}_i \right)
\right. \nn\\
& & \hskip3truecm \left. 
+~\left( 2{A}_i + 2{PA}_i + {PA}_i^s + {PE}_i \right)
\right] ~, \nn\\
P^{\pi^0\pi^0} &=& -  \lambda_t^{(d)} c_i\left[ \left( 2 {PF}_i +
{PF}_i^s + {EX}_i + {\wt EM}_{{\sss C} i} + {P}_i \right) 
\right. \nn\\
& & \hskip3truecm \left. 
+~\left(
2{A}_i + 2{PA}_i + {PA}_i^s + {PE}_i \right) \right]
~.
\label{amplpingpipi}
\eea
In the above, the absence of a prime on the contractions indicates a
$\btod$ transition. Note that, in all cases, the triangle isospin
relation for $\btopipi$ decays is respected:
\beq
\sqrt{2} A^{\pi^-\pi^0} = A^{\pi^-\pi^-} + \sqrt{2} A^{\pi^0\pi^0} ~.
\eeq

In terms of diagrams, the amplitudes for the three $\btopipi$ decays
are given by
\bea
A^{\pi^-\pi^0} &=& -{1\over\sqrt{2}} \left[ (T + C) + (\pew + \pewc)
\right] ~,\nn\\
A^{\pi^+\pi^-} &=& - \left[ ( T + P_u + E + PA_u ) + (P_c + PA_c)
 \right. \nn\\
& & \hskip0.8truecm \left. +~(P_t + PA_t + \frac23 \pewc + \frac13 \pewa -
\frac13\pewe - \frac13\pewpu - \frac13 \pewpau ) \right] ~,\nn\\
A^{\pi^0\pi^0} &=& - {1\over\sqrt{2}} \left[ ( C - P_u - E - PA_u ) +
(-P_c - PA_c ) \right. \\
& & \hskip0.8truecm \left.+~(-P_t - PA_t + \pew + \frac13 \pewc
- \frac13 \pewa + \frac13\pewe + \frac13\pewpu + \frac13 \pewpau )
\right] ~. \nn
\eea
Comparing the amplitudes in terms of contractions and diagrams, we see
that the missing diagrams are given by
\bea
E &=& \sqrt{2} \lambda_u^{(d)} c_i EX^u_i~,\nn\\
PA_u &=& \sqrt{2} \lambda_u^{(d)} c_i PE^u_i~,\nn\\
PA_c &=& \sqrt{2} \lambda_c^{(d)} c_i PE^c_i~,\nn\\ 
PA_t & = & \sqrt{2} \lambda_t^{(d)} c_i \left( 2{A}_i + 2{
PA}_i + {PA}_i^s + {PE}_i \right) ~, \nn\\
\pewa &=& - \sqrt{2} \, \frac{3}{2} \lambda_t^{(d)} c_i { A}_i~,\nn\\
\pewpau&=& -\sqrt{2} \, \frac{3}{2} \lambda_t^{(d)} c_i \left[ ({PA}_i^s
- {PA}_i) + {PE}^s_i \right] ~.
\eea
For diagrams which appear in both $\btopik$ and $\btopipi$ decays, the
expressions for the connection to contractions are the same as in
Eqs.~(\ref{treediagcont}), (\ref{ewpdiagcont}) and
(\ref{pingdiagcont}), except that the prime is removed.

\begin{table} 
\caption{Connection between diagrams and contractions. For the tree
diagrams ($T$, $C$, $P_u$, $P_c$, $A$, $E$, $PA_u$, $PA_c$), a sum
over $i=1,2$ is understood; for the EWP diagrams ($\pew$, $\pewc$,
$\pewpu$, $\pewa$, $\pewe$, $\pewpau$), the sum is over $i=9,10$; for
the gluonic-penguin diagrams ($P_t$, $PA_t$), $i=3$, 4, 5, 6. $q$ is
$d$ or $s$, depending on the $b$ decay.}  \center
\begin{tabular}{ c | c || c | c }
Diagram & Contraction & Diagram & Contraction \\ \hline
$T$ & $\sqrt{2} \lambda^{(q)}_u c_i EM^u_i$ & $\pew$ & $-\sqrt{2} \,
\frac{3}{2} \lambda^{(q)}_t c_i {\wt {EM}}_i$ \\
$C$ & $\sqrt{2} \lambda^{(q)}_u c_i EM^u_{{\sss C} i}$ & $\pewc$ & $-\sqrt{2} \,
\frac{3}{2} \lambda^{(q)}_t c_i {\wt {EM}}_{{\sss C} i}$ \\
$P_u$ & $\sqrt{2} \lambda^{(q)}_u c_i P^u_i$ & $\pewpu$ & $-\sqrt{2} \,
\frac{3}{2} \lambda^{(q)}_t c_i \left[ \left( {PF}^s_i - {PF}_i \right) +
{P}^s_i \right]$ \\
$P_c$ & $-\sqrt{2} \lambda^{(q)}_c c_i P_i^c$ & & \\
$A$ & $\sqrt{2} \lambda^{(q)}_u c_i A^u_i$ & $\pewa$ & $- \sqrt{2} \,
\frac{3}{2} \lambda^{(q)}_t c_i {A}_i$ \\
$E$ & $\sqrt{2} \lambda^{(q)}_u c_i EX^u_i$ & $\pewe$ & $-\sqrt{2} \,
\frac{3}{2} \lambda^{(q)}_t c_i {EX}_i$ \\
$PA_u$ & $\sqrt{2} \lambda^{(q)}_u c_i PE^u_i$ & $\pewpau$ & $-\sqrt{2} \,
\frac{3}{2} \lambda^{(q)}_t c_i \left[ ({PA}_i^s - {PA}_i) + {PE}^s_i \right]$ \\
$PA_c$ & $-\sqrt{2} \lambda^{(q)}_c c_i PE^c_i$ & & \\
\hline
\end{tabular} 
\begin{tabular}{ c | c  }
Diagram & Contraction \\ \hline
$P_t$ & $-\sqrt{2} \lambda_t^{(q)} c_i \left[ 2 {PF}_i +
{PF}_i^s + {EX}_i + {\wt{EM}}_{{\sss C} i} + {P}_i^s \right]$ \\
$PA_t$ & $-\sqrt{2} \lambda_t^{(q)} c_i \left[ 2{A}_i + 2{PA}_i + {PA}_i^s + {PE}_i \right]$ \\
\hline\hline
\end{tabular}
\label{diagcontable}
\end{table}

The connection between diagrams and contractions is summarized in
Table~\ref{diagcontable}. One sees that the expressions for $P_t$ and
$PA_t$ are quite different from those for $P_{u,c}$ and $PA_{u,c}$,
respectively. This is because $P_t$ and $PA_t$ are derived from
contractions involving $O_3$-$O_6$, while $P_{u,c}$ and $PA_{u,c}$
involve $O_1$ and $O_2$.

Note that this connection involves only 8 of the 14 independent
contractions: $EM$, $EM_{{\sss C}}$, $PF$, $P$, $A$, $EX$, $PA$,
$PE$. The six OZI-suppressed contractions $P_{\sss OZI}$, $PF_{\sss
OZI}$, $A_{\sss OZI}$, $PA_{\sss OZI}$, $EX_{\sss OZI}$, $PE_{\sss
OZI}$ are not used. The reason is the following. Each of these six
involves the contraction of ${\bar q}_1$ and $q_2$ and/or ${\bar q}_3$
and $q_4$. This can only happen if the final state contains neutral
particles of the form ${\bar q}q$. In the decays we have considered,
only $\pi^0$ obeys this criterion. However, since $\pi^0 = ({\bar d}d
- {\bar u}u)/\sqrt{2}$, and since we have assumed isospin symmetry,
the six contractions vanish. It is only if we consider final states
involving a $\phi$ (${\bar s}s$) or $J/\Psi$ (${\bar c}c$) that these
contractions will contribute.

\section{Connection to GPY}

In this section, we show that, using contractions, one can reproduce
the main results of Ref.~\cite{GPY}, by Gronau, Pirjol and Yan (GPY).
GPY use a group-theoretical formalism to analyze $\btopik$ decays. On
the other hand, the analysis using contractions is more direct.
Throughout Ref.~\cite{GPY}, GPY neglect the Wilson coefficients (WC's)
$C_7$ and $C_8$, so that the EWP operators are purely
$(V-A)\times(V-A)$. In what follows, we will make the same
approximation. GPY also assume flavour SU(3) symmetry. We will
(eventually) do likewise, but we will show how this assumption is
necessary in our approach.

The first result is the following. The WC's obey $c_1/c_2 =
c_9/c_{10}$ to about 5\%. In the limit in which this equality is
exact, GPY note that the EWP amplitudes ($\pewp$, $\pewcp$, $\pewpup$,
$\pewap$, $\pewep$, $\pewpaup$) are proportional to the tree operators
($T'$, $C'$, $P'_u$, $A'$, $E'$, $PA'_u$).

To see how these relations emerge in our formalism, we first
concentrate on the diagrams $T'$ and $\pewp$ in $\btopik$ decays.
{}From Table \ref{diagcontable}, we see that these are proportional to
$EM$ and ${\wt{EM}}$, respectively. These imply that the final states
$\pi K$ and $K\pi$, respectively, are produced. Now, we note that for
$\pewp$ only the up-quark piece of the electroweak operators
contribute -- this is denoted with an index $u$. That is, $\pewp$
involves only $O_{9,10}^u \sim \bar{s}b \bar{u}u$ (as usual, the
colour and Dirac indices have been suppressed). However, in the SU(3)
limit, $s$-quark contractions are equal to $u$-quark
contractions. Thus, $O_9^u \sim (\bar s_\alpha b_\alpha)_{V-A} (\bar
u_\beta u_\beta)_{V-A} = (\bar u_\alpha b_\alpha)_{V-A}\, (\bar
s_\beta u_\beta)_{V-A} \sim O_1$. Equivalently, $\pi \equiv K$, so
that the final states $\pi K$ and $K\pi$ are the same. Thus,
${\wt{EM}}'_9 = EM'_1$. Things are similar for $O_{10}$ and $O_2$, so
that ${\wt{EM}}'_{10} = EM'_2$. We therefore see that
\beq
T' = \sqrt{2} \lambda_u^{(s)} ( c_1 EM'_1 + c_2 EM'_2 ) = 
\sqrt{2} \lambda_u^{(s)} c_1 \left( EM'_1 + {c_2 \over c_1} EM'_2 \right) ~,
\eeq
while
\beq
\pewp =  -\sqrt{2}\frac{3}{2} \lambda_t^{(s)} ( c_9 {\wt{EM}}'_9 + c_{10} {\wt{EM}}'_{10} ) 
~{\stackrel{SU(3)}{=}}~
 -\sqrt{2}\frac{3}{2} \lambda_t^{(s)}  c_9 \left( EM'_1 + {c_{10} \over c_9} EM'_2 \right) ~.
\eeq
In the limit where $c_1/c_2 = c_9/c_{10}$, $\pewp$ is proportional to $T'$.

The argument is much the same for $C'$ and $\pewcp$. In the SU(3)
limit, we have ${\wt{EM}}'_{{\sss C}9} = EM'_{{\sss C}1}$ and
${\wt{EM}}'_{{\sss C}10} = EM'_{{\sss C}2}$. Thus
\beq
C' = \sqrt{2} \lambda_u^{(s)} ( c_1 EM'_{{\sss C}1} + c_2 EM'_{{\sss C}2} ) = 
\sqrt{2} \lambda_u^{(s)} c_1 \left( EM'_{{\sss C}1} + {c_2 \over c_1} EM'_{{\sss C}2} \right) ~,
\eeq
and
\beq
\pewcp =  -\sqrt{2}\frac{3}{2} \lambda_t^{(s)} ( c_9 {\wt{EM}}'_{{\sss C}9} + c_{10} {\wt{EM}}'_{{\sss C}10} ) 
~{\stackrel{SU(3)}{=}}~
 -\sqrt{2}\frac{3}{2} \lambda_t^{(s)}  c_9 \left( EM'_{{\sss C}1} + {c_{10} \over c_9} EM'_{{\sss C}2} \right) ~.
\eeq
As above, $\pewcp$ is proportional to $C'$ in the limit where $c_1/c_2
= c_9/c_{10}$.

Things are simpler for the other four EWP diagrams $\pewpup$,
$\pewap$, $\pewep$, and $\pewpaup$. In the SU(3) limit, ${PF}^{\prime
s}_i - {PF}'_i = 0$, so that $\pewpup$ is just proportional to $P'_i$
($i=9,10$). Similarly, ${PA}_i^{\prime s} - {PA}'_i = 0$ in the SU(3)
limit, and $\pewpau$ is proportional to $PE'_i$. Thus, all four EWP
diagrams involve the same contractions as the corresponding tree
diagrams $P'_u$, $A'$, $E'$ and $PA'_u$. For $P'_u$ and $\pewpup$ we
can therefore write
\beq
P'_u = \sqrt{2} \lambda_u^{(s)} ( c_1 P'_1 + c_2 P'_2 ) = 
\sqrt{2} \lambda_u^{(s)} c_1 \left( P'_1 + {c_2 \over c_1} P'_2 \right) ~,
\eeq
and
\beq
\pewpup =  -\sqrt{2}\frac{3}{2} \lambda_t^{(s)} ( c_9 P'_1 + c_{10} P'_2 ) =
 -\sqrt{2}\frac{3}{2} \lambda_t^{(s)}  c_9 \left( P'_1 + {c_{10} \over c_9} P'_2 \right) ~.
\eeq
In the limit where $c_1/c_2 = c_9/c_{10}$, $\pewpup$ is proportional
to $P'_u$. The argument is identical for the other three EWP diagrams
$\pewap$, $\pewep$, $\pewpaup$ and the tree diagrams $A'$, $E'$,
$PA'_u$.

We therefore see that the proportionality constant between the EWP and
corresponding tree diagrams is $- (3/2) (c_9/c_1) \lambda^{(q)}_t /
\lambda^{(q)}_u$, which agrees with GPY. In particular, we have
\beq
\frac{\pewp}{T'} = -\frac{3}{2} \frac{\lambda_t^{(s)}}{\lambda_u^{(s)}}
\frac{c_9}{c_1}~,~~~~~\frac{\pewcp}{C'} = -\frac{3}{2}
\frac{\lambda_t^{(s)}}{\lambda_u^{(s)}} \frac{c_{9}}{c_1}~.
\label{eq2}
\eeq

The second result is two relations between EWP and tree diagrams:
\bea
\label{GPYrels}
P_{\sss EW}(B^- \to \pi ^- \bar K ^0) + \sqrt{2} P_{\sss EW}(B^- \to
\pi ^0 \bar K ^-) &=&\frac{3}{2}\frac{c_9+c_{10}}{c_1+c_2}R(C'+T')~,\nn\\
P_{\sss EW}(\bar B^0 \to \pi ^+ K ^-) + P_{\sss EW}(B^- \to \pi ^-
\bar K ^0) &=& \nn\\\
& & \hskip-7truecm \frac{3}{4}\frac{c_9-c_{10}}{c_1-c_2}R(A'+C'-T'-E)
- \frac{3}{4}\frac{c_9+c_{10}}{c_1+c_2}R(A'-C'-T'+E)~,
\eea
where $R= \lambda^{(s)}_t /\lambda^{(s)}_u$.  All the diagrams in the
above relations are for $\btopik$ decays, with the exception of $E$,
which is for $\btopipi$. However, since flavour SU(3) has been
assumed, $\btopik$ and $\btopipi$ diagrams are equal.

In order to reproduce these relations, we make the following
observation. In the limit of neglecting $C_7$ and $C_8$, both EWP and
tree operators are $(V-A)\times(V-A)$. In this case, it is possible to
do a Fierz transformation to exchange the position of $q_5$ and $q_6$
without changing the Dirac structure of the operators. This results in
$O_1 \leftrightarrow O_{10}$ and $O_2 \leftrightarrow O_9$. If we also
take the SU(3) limit, in which case one can switch $u$ and $s$ quarks,
we have $O_1 = O_9$ and $O_2 = O_{10}$. Given the equivalence of
different operators, this implies that certain contractions are
pairwise the same (within each of the four contraction classes):
\bea
EM_{1,9} \equiv EM_{{\sss C} 2,10} & ~,~~~ & 
EM_{2,10} \equiv EM_{{\sss C} 1,9} ~, \nn\\
P_{1,9} \equiv PF_{2,10} & ~,~~~ &
P_{2,10} \equiv PF_{1,9} ~, \nn\\
P_{{\sss OZI} {1,9}} \equiv PF_{{\sss OZI} 2,10} & ~,~~~ & 
P_{{\sss OZI} 2,10} \equiv PF_{{\sss OZI} {1,9}} ~, \nn\\
A_{1,9} \equiv EX_{2,10} & ~,~~~ & 
A_{2,10} \equiv EX_{1,9} ~, \nn\\
A_{{\sss OZI} 1,9} \equiv EX_{{\sss OZI} 2,10} & ~,~~~ & 
A_{{\sss OZI} 2,10} \equiv EX_{{\sss OZI} 1,9} ~, \nn\\
PA_{1,9} \equiv PE_{2,10} & ~,~~~ & 
PA_{2,10} \equiv PE_{1,9} ~, \nn\\
PA_{{\sss OZI} 1,9} \equiv PE_{{\sss OZI} 2,10} & ~,~~~ &
PA_{{\sss OZI} 2,10} \equiv PE_{{\sss OZI} 1,9} ~.
\eea

Consider now the first relation.  The left-hand side of the equality
is
\bea
\sum_{i=9,10} \sqrt{2} \frac{3}{2} \lambda^{(s)}_t c_i \left( {\wt
{EM}}'_i + {\wt {EM}}'_{{\sss C} i} \right) & & \nn\\
& & \hskip-1truein
=~\sqrt{2} \frac{3}{2} \lambda^{(s)}_t \left( c_9 {\wt {EM}}'_9 + c_{10} {\wt {EM}}'_{10}
+ c_9 {\wt {EM}}'_{{\sss C}9}+ c_{10} {\wt {EM}}'_{{\sss C}10} \right)\nn\\
& & \hskip-1.1truein
\stackrel{SU(3)}{=}
~\sqrt{2} \frac{3}{2} \lambda^{(s)}_t \left( c_9 EM'_1 + c_{10} EM'_2 +
c_9 EM'_{{\sss C}1} + c_{10} EM'_{{\sss C}2} \right) \nn\\
& & \hskip-1.3truein 
\stackrel{Fierz+SU(3)}{=}
~\sqrt{2} \frac{3}{2} \lambda^{(s)}_t \left( c_9 EM'_1 + c_{10} EM'_2 + c_9 EM'_2
+ c_{10} EM'_1 \right) \nn\\
& & \hskip-1truein
= ~\sqrt{2} \frac{3}{2}\lambda^{(s)}_t (c_9 +c_{10})( EM'_1 + EM'_2 ) ~.
\eea
The right-hand side is
\bea
\sum_{i=1,2} \sqrt{2} \frac{3}{2}\lambda^{(s)}_t\frac{c_9+c_{10}}{c_1+c_2} (c_i EM'_i + c_i
EM'_{{\sss C} i} ) & & \nn\\
& & \hskip-1truein
=~\sqrt{2} \frac{3}{2}\lambda^{(s)}_t\frac{c_9+c_{10}}{c_1+c_2} (c_1 EM'_1 + c_2 EM'_2
+ c_1 EM'_{{\sss C}1} + c_2 EM'_{{\sss C}2}) \nn\\
& & \hskip-1.3truein 
\stackrel{Fierz+SU(3)}{=}
~\sqrt{2} \frac{3}{2}\lambda^{(s)}_t\frac{c_9+c_{10}}{c_1+c_2} (c_1 EM'_1 + c_2
EM'_2 + c_1 EM'_2 + c_2 EM'_1) \nn\\
& & \hskip-1truein
=~\sqrt{2} \frac{3}{2}\lambda^{(s)}_t\frac{c_9+c_{10}}{c_1+c_2} \left[ (c_1 +
c_2) EM'_1 + (c_1 + c_2) EM'_2 \right] \nn\\
& & \hskip-1truein
=~\sqrt{2} \frac{3}{2}\lambda^{(s)}_t(c_9+c_{10}) (EM'_1 + EM'_2 ) ~.
\eea
We therefore reproduce the first GPY EWP-tree relation using
contractions.

We now turn to the second relation of GPY. The left-hand side of the
equality is
\bea
\sum_{i=9,10} c_i \sqrt{2} \frac{3}{2} \lambda^{(s)}_t\left( {\wt {EM}}'_{{\sss C} i}
- EX'_i \right) &=&\sqrt{2} \frac{3}{2} \lambda^{(s)}_t\left( c_9 {\wt {EM}}'_{{\sss
    C} 9} + c_{10} {\wt{EM}}'_{{\sss C} 10} - c_9 EX'_9 - c_{10}
EX'_{10}\right)\nn\\
&=&\sqrt{2} \frac{3}{2} \lambda^{(s)}_t\left(c_9 EM'_2 +c_{10} EM'_1 - c_9 EX'_1 -
c_{10} EX'_2 \right) ~,
\eea
and the right-hand side of the equality is
\bea
&& \hskip-1.3truecm \sum_{i=1,2}
\sqrt{2}\frac{3}{4} \lambda^{(s)}_t\left[ c_i\frac{c_9-c_{10}}{c_1-c_2}(A'_i+EM'_{{\sss
C}i}-EM'_i-EX'_i) \right. \nn\\
& & \hskip0.6truein \left. -~c_i\frac{c_9+c_{10}}{c_1+c_2}(A'_i-EM'_{{\sss
C}i}-EM'_i+EX'_i) \right] \nn\\
&=&\sqrt{2}\frac{3}{4}\lambda^{(s)}_t\frac{c_9-c_{10}}{c_1-c_2} \left( c_1 A'_1+ c_2
A'_2+ c_1 EM'_{{\sss C}1} +c_2 EM'_{{\sss C}2} \right. \nn\\
& & \hskip1truein
\left. - c_1 EM'_1 - c_2 EM'_2 - c_1 EX'_1 -c_2 EX'_2 \right) \nn\\
&&- \sqrt{2}\frac{3}{4}\lambda^{(s)}_t\frac{ c_9 + c_{10}}{c_1 + c_2} \left( c_1
A'_1+ c_2 A'_2- c_1 EM'_{{\sss C}1} -c_2 EM'_{{\sss C}2} \right. \nn\\
& & \hskip1.2truein
\left. - c_1 EM'_1 - c_2 EM'_2 + c_1 EX'_1 +c_2 EX'_2 \right) \nn\\
&=&\sqrt{2}\frac{3}{4}\lambda^{(s)}_t\frac{c_9-c_{10}}{c_1-c_2} \left( c_1 EX'_2+ c_2
EX'_1+ c_1 EM'_2 +c_2 EM'_1 \right. \nn\\
& & \hskip1truein \left.  - c_1 EM'_1 - c_2 EM'_2 - c_1 EX'_1 -c_2
EX'_2 \right)\nn\\
&&- \sqrt{2}\frac{3}{4}\lambda^{(s)}_t\frac{ c_9 + c_{10}}{c_1 + c_2} \left( c_1
EX'_2+ c_2 EX'_1- c_1 EM'_2 -c_2 EM'_1 \right. \nn\\
& & \hskip1.2truein \left. - c_1 EM'_1 - c_2 EM'_2 + c_1 EX'_1 +c_2
EX'_2 \right)\nn\\
&=&\sqrt{2}\frac{3}{4}\lambda^{(s)}_t\frac{c_9-c_{10}}{c_1-c_2}(c_1-c_2)
(EX'_2-EX'_1+EM'_2-EM'_1)\nn\\
&& \hskip0.6truein -~\sqrt{2}\frac{3}{4}\lambda^{(s)}_t\frac{ c_9 + c_{10}}{c_1 +
 c_2}(c_1+c_2)(EX'_2+EX'_1-EM'_2-EM'_1)\nn\\
&=&\sqrt{2} \frac{3}{2} \lambda^{(s)}_t\left(c_9 EM'_2 +c_{10} EM'_1 - c_9 EX'_1 -
 c_{10} EX'_2\right)~.
\eea
This proves the second EWP-tree GPY relation in $\btopik$.

We have therefore shown explicitly that, using contractions, we
reproduce the main results of GPY \cite{GPY} for $\btopik$ decays.

It is worth making one final remark here. We have mentioned in
Sec.~2.1 that the contractions $A'_i$ and $EX'_i$ are higher order in
$(1/m_b)$. If these contributions are indeed small, then the second
GPY relation can be separated into two relations, one leading order
(LO), the other next-to-leading order (NLO). These contain the large
and small contributions, respectively. We have
\bea
\left[ P_{\sss EW}(\bar B^0 \to \pi ^+ K ^-) + P_{\sss EW}(B^- \to \pi ^-
\bar K ^0) \right]_{\sss LO}&=& \nn\\
& & \hskip-7truecm \frac{3}{4}\frac{c_9-c_{10}}{c_1-c_2}R(C'-T')
- \frac{3}{4}\frac{c_9+c_{10}}{c_1+c_2}R(C'-T')~, \nn\\
\left[ P_{\sss EW}(\bar B^0 \to \pi ^+ K ^-) + P_{\sss EW}(B^- \to \pi ^-
\bar K ^0)\right]_{\sss NLO} &=& \nn\\
& & \hskip-7truecm \frac{3}{4}\frac{c_9-c_{10}}{c_1-c_2}R(A'-E)
- \frac{3}{4}\frac{c_9+c_{10}}{c_1+c_2}R(A'+E)~.
\eea
Of course, each of these relations is individually satisfied.

\section{Connection to QCDfac and pQCD}

All the results in the previous sections hold to all orders in
$\alpha_s$. As mentioned earlier, in our general approach, we are
interested only in the possible ways in which the final-state quarks,
which evolve into the final-state mesons, can be produced in $B$
decays through the effective Hamiltonian. That is, we do not attempt
to calculate the strong interactions of the final-state quarks through
the exchange of gluons and their ultimate hadronization into the
final-state mesons. However, it is also possible to work
order-by-order in $\alpha_s$ by including the gluons explicitly.  This
is useful as it allows us to make a connection with QCDfac and pQCD,
both of which work order-by-order. They do this because in their
frameworks the nonleptonic amplitudes are expanded in powers of
$\alpha_s(\mu_h)$ ($\mu_h \sim \sqrt{\Lambda_{\sss QCD}m_b}$) for
gluon exchanges involving the spectator quark, or $\alpha_s(m_b)$ for
the remaining gluon exchanges.

The first calculation involves gluon exchanges only between quarks
that belong to the same meson. In our approach, such gluon-exchange
effects are taken into account automatically, and so we call this the
$O(\alpha_s^0)$ (``zero-order'') calculation. This is in fact just the
factorization assumption in calculating nonleptonic amplitudes.

The exchanged gluon between the quarks in a meson can be soft and
hence these contributions are not perturbatively calculable. In the
pQCD approach, the soft gluon-exchange contributions involving the
soft spectator quark in the $B$ meson are argued to be highly
suppressed.  Thus, the soft spectator quark has to be boosted up in
energy through gluon exchange in order to end up in the final-state
light meson with energy $E \sim M_{\sss B}$. Since the gluon is
relatively hard such contributions are calculable perturbatively in
pQCD. However, it is not entirely clear that soft-gluon exchanges
involving the spectator quark can be neglected, and in the QCDfac
approach the soft-gluon contributions are absorbed in physical form
factors.

In the second step we consider gluons exchanged between quarks
belonging to different mesons [$O(\alpha_s^1)$]. In our approach we do
not calculate the amplitudes explicitly. Instead, we work out the
general structure of the contractions when one-gluon exchanges are
included among the quarks that emerge from the $B$ decay through the
weak effective Hamiltonian.

In the pQCD approach it is argued that such corrections are calculable
perturbatively, while in the QCDfac approach the gluon exchanges
involving the spectator quark are sensitive to long-distance physics.
Thus, their effects, like the form factors, are represented by
additional unknown hadronic quantities that may be obtained from a fit
to data.

\subsection{Order zero}

In this section we will consider the zero-order calculation. From here
on, we concentrate only on $T'$, $C'$, $\pewp$ and $\pewcp$ in
$\btopik$ decays. In this case, with no gluons between quarks of
different mesons, it is possible to take colours into account simply
by counting them.  For example, for $T$ we have
\bea
\frac{T}{\sqrt{2}} & = & \sum_{i=1,2} c_i EM_i \nn\\
&=&
c_1
{
\contraction[1ex]{\langle \bar q_{1x} q_{2x} \bar q_{3y} }{q_{4y} }{| \bar q_{5\alpha} b_\alpha }{\bar q_{6\beta} }
\contraction[2ex]{\langle \bar q_{1x} q_{2x} }{\bar q_{3y} }{q_{4y} | \bar q_{5\alpha} b_\alpha \bar q_{6\beta} }{q_{7\beta} }
\contraction[3ex]{\langle \bar q_{1x} }{q_{2x} }{\bar q_{3y} q_{4y} | }{\bar q_{5\alpha} }
\contraction[4ex]{\langle }{\bar q_{1x} }{q_{2x} \bar q_{3y} q_{4y} | \bar q_{5\alpha} b_\alpha \bar q_{6\beta} q_{7\beta} | }{\bar q_{8z}}
\langle \bar q_{1x} q_{2x} \bar q_{3y} q_{4y} | \bar q_{5\alpha} b_\alpha \bar q_{6\beta} q_{7\beta} | \bar q_{8z} b_z \rangle
}
+
c_2
{
\contraction[1ex]{\langle \bar q_{1x} q_{2x} \bar q_{3y} }{q_{4y} }{| \bar q_{5\alpha} b_\beta }{\bar q_{6\beta} }
\contraction[2ex]{\langle \bar q_{1x} q_{2x} }{\bar q_{3y} }{q_{4y} | \bar q_{5\alpha} b_\beta \bar q_{6\beta} }{q_{7\beta} }
\contraction[3ex]{\langle \bar q_{1x} }{q_{2x} }{\bar q_{3y} q_{4y} | }{\bar q_{5\alpha} }
\contraction[4ex]{\langle }{\bar q_{1x} }{q_{2x} \bar q_{3y} q_{4y} | \bar q_{5\alpha} b_\beta \bar q_{6\beta} q_{7\alpha} | }{\bar q_{8z}}
\langle \bar q_{1x} q_{2x} \bar q_{3y} q_{4y} | \bar q_{5\alpha} b_\beta \bar q_{6\beta} q_{7\alpha} | \bar q_{8z} b_z \rangle
}~.
\eea
In the above, the subscripts $\alpha$, $\beta$, $x$, $y$ and $z$ are
colour indices. Since gluon exchanges among quarks belonging to
different mesons are neglected, the contractions are simply
proportional to delta functions in the colours:
\bea
\sum_{i=1,2} c_i EM_i
&=&
c_1
\delta_{xz}\delta_{x\alpha}\delta_{y\beta}\delta_{y\beta}\delta_{\alpha
  z}
{
\contraction[1ex]{\langle \bar q_1 q_2 \bar q_3 }{q_4 }{| \bar q_5 b }{\bar q_6 }
\contraction[2ex]{\langle \bar q_1 q_2 }{\bar q_3 }{q_4 | \bar q_5 b \bar q_6 }{q_7 }
\contraction[3ex]{\langle \bar q_1 }{q_2 }{\bar q_3 q_4 | }{\bar q_5 }
\contraction[4ex]{\langle }{\bar q_1 }{q_2 \bar q_3 q_4 | \bar q_5 b \bar q_6 q_7 | }{\bar q_8}
\langle \bar q_1 q_2 \bar q_3 q_4 | \bar q_5 b \bar q_6 q_7 | \bar q_8 b \rangle
}
+
c_2
\delta_{xz}\delta_{x\alpha}\delta_{y\alpha}\delta_{y\beta}\delta_{\beta
  z}
{
\contraction[1ex]{\langle \bar q_1 q_2 \bar q_3 }{q_4 }{| \bar q_5 b }{\bar q_6 }
\contraction[2ex]{\langle \bar q_1 q_2 }{\bar q_3 }{q_4 | \bar q_5 b \bar q_6 }{q_7 }
\contraction[3ex]{\langle \bar q_1 }{q_2 }{\bar q_3 q_4 | }{\bar q_5 }
\contraction[4ex]{\langle }{\bar q_1 }{q_2 \bar q_3 q_4 | \bar q_5 b \bar q_6 q_7 | }{\bar q_8}
\langle \bar q_1 q_2 \bar q_3 q_4 | \bar q_5 b \bar q_6 q_7 | \bar q_8 b \rangle
}\nn\\
&=&
c_1 N_c^2
{
\contraction[1ex]{\langle \bar q_1 q_2 \bar q_3 }{q_4 }{| \bar q_5 b }{\bar q_6 }
\contraction[2ex]{\langle \bar q_1 q_2 }{\bar q_3 }{q_4 | \bar q_5 b \bar q_6 }{q_7 }
\contraction[3ex]{\langle \bar q_1 }{q_2 }{\bar q_3 q_4 | }{\bar q_5 }
\contraction[4ex]{\langle }{\bar q_1 }{q_2 \bar q_3 q_4 | \bar q_5 b \bar q_6 q_7 | }{\bar q_8}
\langle \bar q_1 q_2 \bar q_3 q_4 | \bar q_5 b \bar q_6 q_7 | \bar q_8 b \rangle
}
+
c_2 N_c
{
\contraction[1ex]{\langle \bar q_1 q_2 \bar q_3 }{q_4 }{| \bar q_5 b }{\bar q_6 }
\contraction[2ex]{\langle \bar q_1 q_2 }{\bar q_3 }{q_4 | \bar q_5 b \bar q_6 }{q_7 }
\contraction[3ex]{\langle \bar q_1 }{q_2 }{\bar q_3 q_4 | }{\bar q_5 }
\contraction[4ex]{\langle }{\bar q_1 }{q_2 \bar q_3 q_4 | \bar q_5 b \bar q_6 q_7 | }{\bar q_8}
\langle \bar q_1 q_2 \bar q_3 q_4 | \bar q_5 b \bar q_6 q_7 | \bar q_8 b \rangle
}\nn\\
&=& c_1 N_c^2 \, {\overline{EM}} + c_2 N_c \, {\overline{EM}} \nn\\
&=& (c_1  + \frac{c_2} {N_c})N_c^2 \, {\overline{EM}} ~,
\eea
where the bar on ${\overline{EM}}$ is added to stress the fact that
colour effects are extracted.  The idea is the following: \textit{a
priori} the matrix elements $EM_1$ and $EM_2$ are different.  But
since the difference between $O_1$ and $O_2$ is only their colour
structure, once we have taken this into account by explicitly adding
the effect of colour, the remaining matrix elements are identical.
Doing this for $T'$, $C'$, $\pewp$ and $\pewcp$, it is easy to derive
the following relations
\bea
T'&=& \sqrt{2} (c_1  + \frac{c_2} {N_c}) N_c^2 \, {\overline{EM}}' ~,\nn\\
C'&=& \sqrt{2} (\frac{c_1} {N_c} + c_2 )N_c^2 \, {\overline{EM}_{\sss C}}'~,\nn\\
\pewp &=&  -\frac{3}{2} \sqrt{2} (c_9  +\frac{c_{10}}{N_c})N_c^2
\, {\overline{{\widetilde {EM}}}}'~,\nn\\
\pewcp &=& - \frac{3}{2} \sqrt{2} (\frac{c_9} {N_c} +c_{10})N_c^2
\, {\overline{{\widetilde {EM}}}}_{\sss C}'~.
\eea
With SU(3), and taking into account Fierz transformations, we have
${\overline{EM}}' = {\overline{EM}}_{\sss C}' = {\overline{{\widetilde
{EM}}}}' = {\overline{{\widetilde {EM}}}}_{\sss C}'$. Then we obtain
the two EWP-tree results of Eq.~(\ref{eq2}) (which hold to
all-orders), and a third relation:
\beq
{C' \over T'} = { (c_1 + c_2 N_c) \over (c_1 N_c + c_2)} = 0.17 \pm 0.1 ~.
\label{CTrel}
\eeq
The numerical value of this ratio is obtained as follows. At leading
order the WC's take the values $c_1 = 1.081$ and $c_2 = -0.190$ (these
values are taken from the last paper in Ref.~\cite{BBNS}). Thus,
taking $N_c = 3$, there is some cancellation between the factors in
the numerator of Eq.~(\ref{CTrel}). Now, the WC's are calculated using
a renormalization point of $\mu = m_b$. However, in fact we do not
know the precise value of $\mu$ -- all we know is that it is $O(m_b)$.
Allowing the values of the WC's to vary leads to an uncertainty on the
above ratio. The central value uses the above values for $c_1$ and
$c_2$, but the (estimated) error corresponds to allowing $\mu$ to
vary. The bottom line is that the precise value of the ratio $C'/T'$
is uncertain. Still, although this ratio can be zero in the case of
complete cancellation, there is an upper limit of approximately $1/3$.
We note in passing that the naive estimates of colour suppression are
supported by the WC calculations.

Above, we have concentrated on $T'$, $C'$, $\pewp$ and $\pewcp$ in
$\btopik$ decays. However, one can perform the above procedure for
other contractions/diagrams. If one does so for the annihilation and
exchange diagrams $A$ and $E$, one finds the following interesting
relation. At $O(\alpha_s^0)$, the matrix elements are equal, so that
\beq
\left\vert \frac{A'}{E'} \right\vert = \frac{c_1
  +\frac{c_2}{N_c}}{\frac{c_1}{N_c}+c_2}\approx 6.0 ~.
\eeq
We therefore see that, at leading order, the diagram $A'$ is expected
to be much larger than $E'$. BS found a similar result, but in the
large $N_c$ limit.

\subsection{Corrections with 1 gluon}

\begin{figure}[ht]
\centerline{\epsfxsize=3.6truein \epsfbox{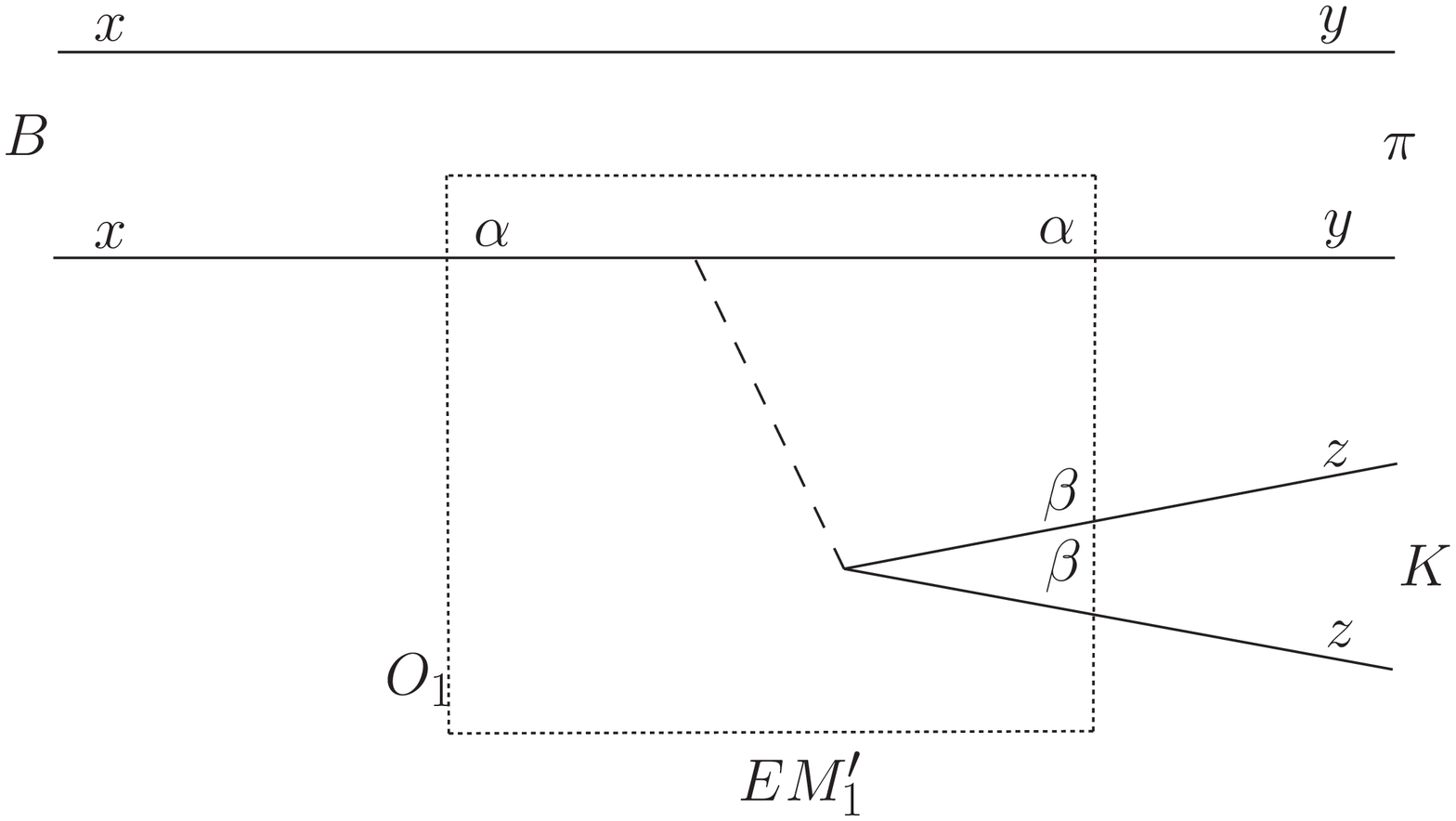}
 ~\epsfxsize=3.6truein \epsfbox{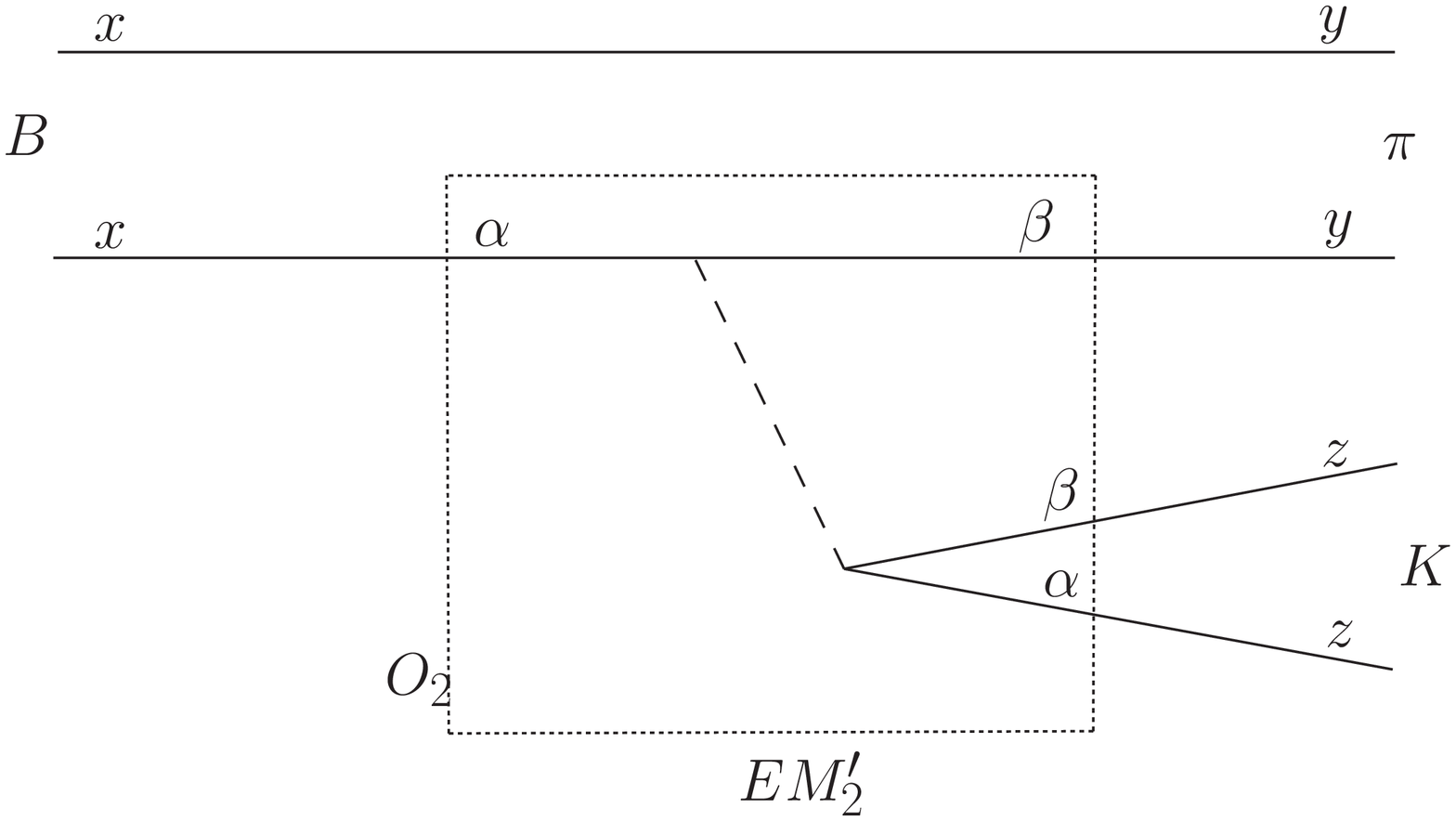}}
\caption{$EM'$ contractions of tree operators for $\btopik$.}
\label{fige}
\end{figure}

\begin{figure}[ht]
\centerline{\epsfxsize=3.6truein \epsfbox{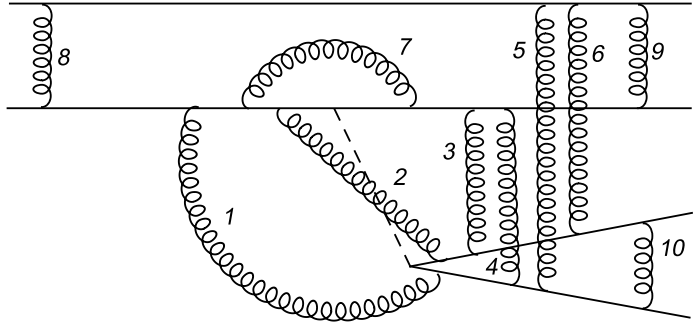}}
\caption{One-gluon corrections to $EM$}
\label{figconte}
\end{figure}

We now turn to the $O(\alpha_s^1)$ QCD corrections. We introduce a
single gluon between all pairs of quarks in each
diagram/contraction. We illustrate this procedure by considering the
$EM$-type contraction for tree operators in $\btopik$
(Fig.~\ref{fige}). There are ten possibilities for the gluon, shown in
Fig.~\ref{figconte}.  Note that these include a possible gluon
exchange between the quarks within the same meson -- we will be able
to absorb these corrections into the order-zero result.

We introduce the following notation: $Z_{i,j}$ is the $j^{\rm th}$ QCD
correction ($j=1,...,10$) to the $Z$-type contraction ($Z=EM$,
$EM_{\sss C}$,..., ) of $O_i$.  $\bar Z_{i,j}$ is the same, except
that the colours have been explicitly extracted: $Z_{i,j} = (colour
factor)\times \bar Z_{i,j}$).

The idea is the same as for the $O(\alpha_s^0)$ effects: the only
difference between $O_1$ and $O_2$ is the colour structure. For
example, $EM_{1,i}$ and $EM_{2,i}$ differ only in their colour
structure, and so their matrix elements are equal once the colour
effects have been extracted. The problem is more complicated here
because there are several QCD corrections, e.g.\ we cannot relate
$EM_{1,1}$ to $EM_{1,2}$. Thus, the overall matrix elements are not
equal unless all contributions appear in the same linear
combination. We will see this explicitly below. Note also that matrix
elements are still not evaluated.

We begin by computing the colour factor for $EM_{1,1}$
\bea
\delta_{yx}\delta_{xm}T^a_{m\alpha}\delta_{\alpha y}\delta_{z\beta}T^a_{\beta n}\delta_{nz}
&=&\delta_{yx}\delta_{xm}\delta_{\alpha y}\delta_{z\beta}\delta_{nz}T^a_{m\alpha}T^a_{\beta n}\nn\\
&=&\delta_{yx}\delta_{xm}\delta_{\alpha y}\delta_{z\beta}\delta_{nz} \left
(-\frac{1}{2N_c} \delta_{m\alpha}\delta_{\beta n} +\frac{1}{2} \delta_{mn}\delta_{\alpha\beta} \right)\nn\\
&=&-\frac{N_c^2}{2N_c} + \frac{N_c}{2}=0~.
\eea
Then $EM_{1,1}$ vanishes by colour arguments.  For $EM_{2,1}$,
\bea
\delta_{yx}\delta_{xm}T^a_{m\alpha}\delta_{\beta y}\delta_{z\beta}T^a_{\alpha n}\delta_{nz}
&=&\delta_{yx}\delta_{xm}\delta_{\beta y}\delta_{z\beta}\delta_{nz}T^a_{m\alpha}T^a_{\alpha n}\nn\\
&=&\delta_{yx}\delta_{xm}\delta_{\beta y}\delta_{z\beta}\delta_{nz}
\left(-\frac{1}{2N_c} \delta_{m\alpha}\delta_{\alpha n} +\frac{1}{2} \delta_{mn}\delta_{\alpha\alpha}\right)\nn\\
&=&-\frac{N_c}{2N_c} + \frac{N_c^2}{2}=\frac{N_c^2}{2}\left( 1-\frac{1}{N_c^2}\right)~.
\eea
The total contribution of the correction 1 to the contraction $EM$ is
then
\beq
c_1 EM_{1,1} + c_2 EM_{2,1}
= c_2 \frac{N_c^2}{2}\left( 1-\frac{1}{N_c^2}\right) {\overline{EM}}_{1}~.
\eeq

A similar calculation has been performed for the remaining nine QCD
corrections of the contraction $EM$, with the following result:
\beq
\sum_{i=1}^{10} (c_1 EM_{1,i} + c_2 EM_{2,i}) = c_2 \frac{N_c^2}{2}\left(
1-\frac{1}{N_c^2}\right) \sum_{i=1}^6 {\overline{EM}}_i + \left(
c_1+\frac{c_2}{N_c}\right) \frac{N_c^3}{2} \left(
1-\frac{1}{N_c^2}\right) \sum_{i=7}^{10} {\overline{EM}}_i~.
\eeq
Because the coefficients of the two terms are not the same, the
overall $O(\alpha_s^1)$ matrix element of $O_1$ is not the same as
that of $O_2$.

We can apply this to the $T'$ diagram in $\btopik$ decays. Let
$T'=T^{\prime 0}+T^{\prime 1}$ where $T^{\prime 0}$ is the piece
without gluons [$O(\alpha_s^0$)] and $T^{\prime 1}$ is the piece with
one-gluon corrections [$O(\alpha_s^1$)]. We have
\bea
T^{\prime 0} &=& \sqrt{2} (c_1  + \frac{c_2} {N_c})N_c^2 \, {\overline{EM}}' ~,\nn\\
T^{\prime 1} &=& \sqrt{2} \left[ c_2 \frac{N_c^2}{2}\left(
1-\frac{1}{N_c^2}\right) \sum_{i=1}^6 {\overline{EM}}_i + \left(
c_1+\frac{c_2}{N_c}\right) \frac{N_c^3}{2} \left(
1-\frac{1}{N_c^2}\right) \sum_{i=7}^{10} {\overline{EM}}_i \right]~.
\label{T_alpha_s}
\eea
However, an examination of Fig.~\ref{figconte} reveals that the
corrections $i=7,8,9,10$ all correspond to gluon exchange between
quarks within the same meson. As such, these corrections can be
absorbed in the $O(\alpha_s^0)$ result. We will therefore rewrite
Eq.~\ref{T_alpha_s} as as
\bea
T^{\prime 0} &=& \sqrt{2} (c_1  + \frac{c_2} {N_c}) N_c^2 \, {\overline{EM}}'_{\sss F} ~~,~~~~
{\overline{EM}}'_{\sss F} \equiv {\overline{EM}}' +
\frac{N_c}{2} \left(
1-\frac{1}{N_c^2}\right) \sum_{i=7}^{10} {\overline{EM}}_i ~, \nn\\
T^{\prime 1} &=& \sqrt{2} \, \frac{c_2}{2} N_c^2\left(
1-\frac{1}{N_c^2}\right) \sum_{i=1}^6 {\overline{EM}}_i ~.
\eea

\begin{figure}[ht]
\centerline{\epsfxsize=3.6truein \epsfbox{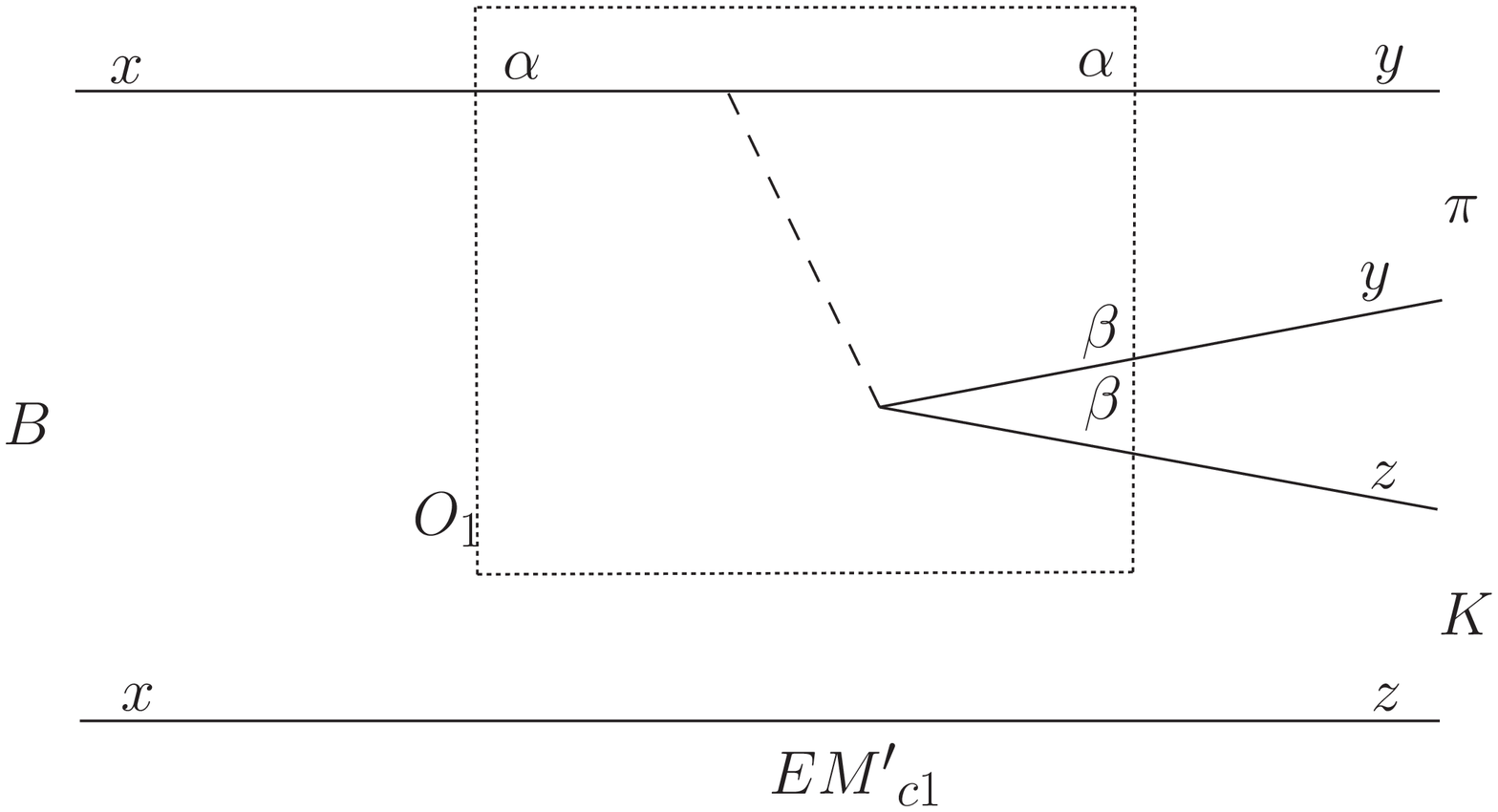} ~\epsfxsize=3.6truein \epsfbox{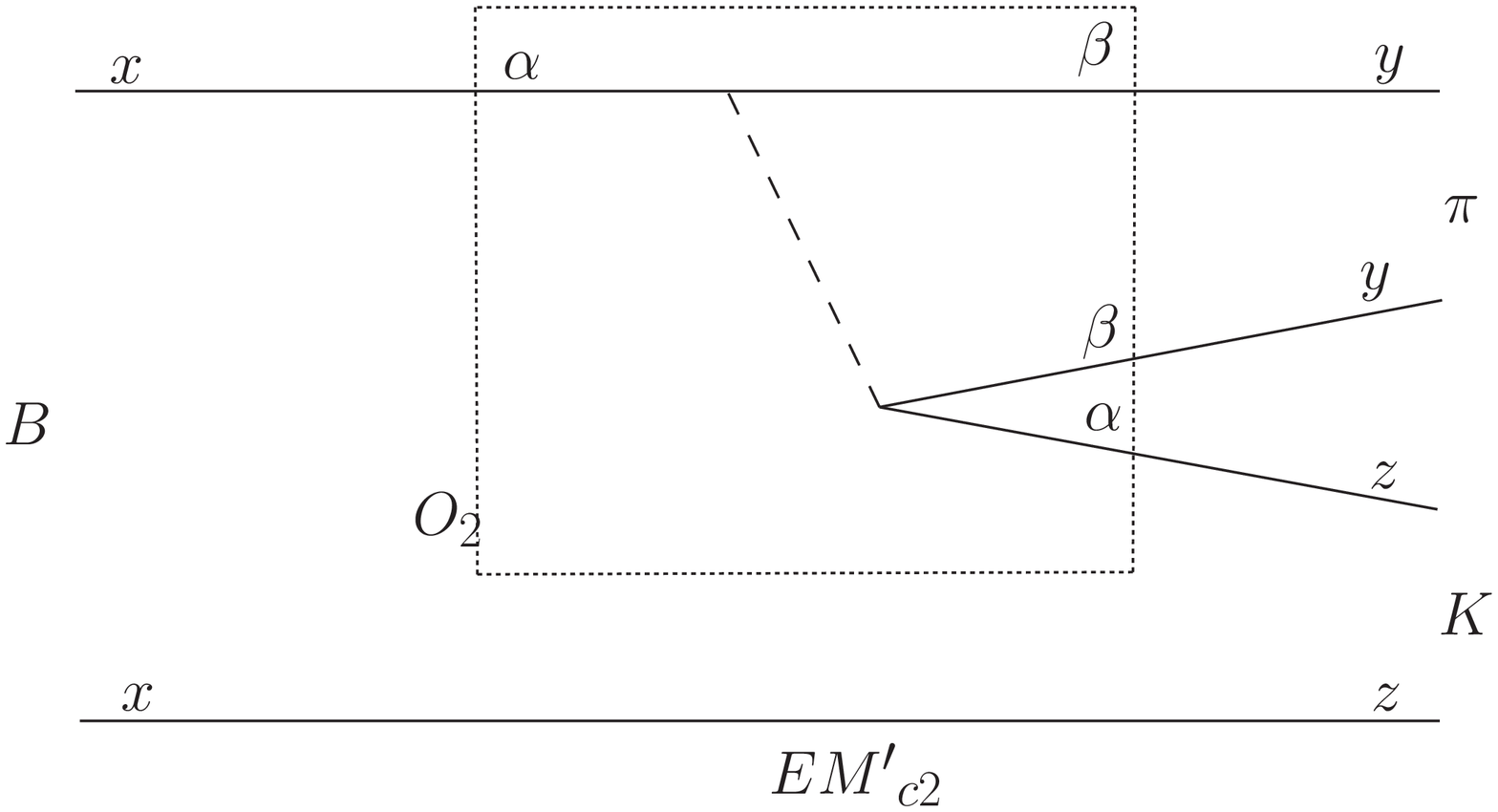}}
\caption{$EM'_{\sss C}$ contractions of colour-suppressed tree operators for $\btopik$.}
\label{figf}
\end{figure}

\begin{figure}[ht]
\centerline{\epsfxsize=3.6truein \epsfbox{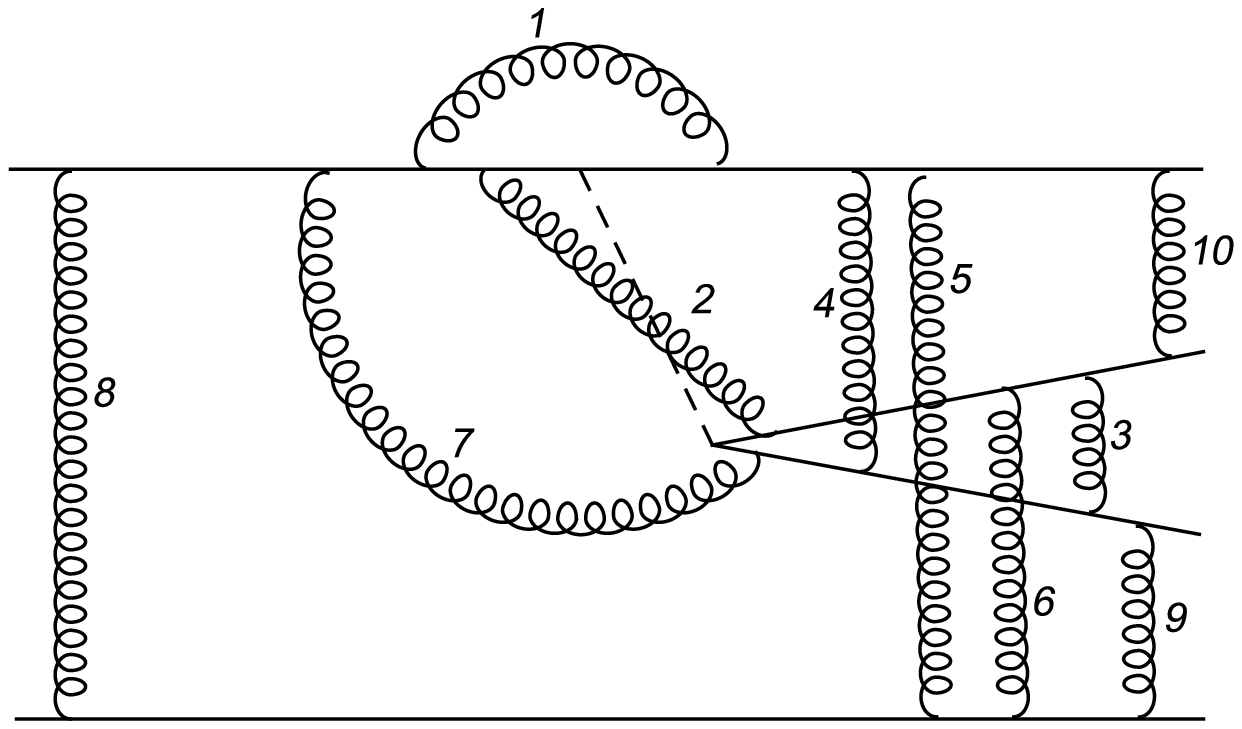}}
\caption{One-gluon corrections to $EM_{\sss C}$}
\label{figcontf}
\end{figure}

This procedure can also be carried out for the $EM_{\sss C}$
contraction.  Fig.~\ref{figf} shows the $EM_{\sss C}$-type contraction
for colour-suppressed tree operators in $\btopik$. As before, we
consider a single gluon exchange between all pairs of quarks. There
are ten placements for these gluons, shown in Fig.~\ref{figcontf}.

Taking all one-gluon QCD corrections into account, we find
\beq
\sum_{i=1}^{10} (c_1 EM_{\sss C1,i} + c_2 EM_{\sss C2,i}) = c_1
\frac{N_c^2}{2}\left( 1-\frac{1}{N_c^2}\right) 
\sum_{i=1}^6 \overline{EM}_{{\sss C}i} 
+ \left(\frac{c_1}{N_c}+ c_2\right) \frac{N_c^3}{2} \left(
1-\frac{1}{N_c^2}\right)
\sum_{i=7}^{10} {\overline{EM}_{{\sss C}i}} \
\eeq
Thus, as for the $EM$ contraction, the overall $O(\alpha_s^1)$ matrix
element of $O_1$ is not the same as that of $O_2$.

We now apply this to the $C'$ diagram in $\btopik$ decays: as with
$T'$ we define $C'=C^{\prime 0}+C^{\prime 1}$, with
\bea
C^{\prime 0} &=& \sqrt{2} (c_1 N_c + c_2 N_c^2) \, {\overline{EM}}_{\sss C}' ~,\nn\\
C^{\prime 1} &=& \sqrt{2} \left[ c_1 \frac{N_c^2}{2}\left(
1-\frac{1}{N_c^2}\right) \sum_{i=1}^6 \overline{EM}_{{\sss C}i} + \left(
\frac{c_1}{N_c}+c_2\right) \frac{N_c^3}{2} \left(
1-\frac{1}{N_c^2}\right) \sum_{i=7}^{10} \overline{EM}_{{\sss C}i} \right]~.\eea
However, like the $T'$ contribution, the corrections $i=7,8,9,10$ can
all be absorbed in the $O(\alpha_s^0)$ result.  We therefore rewrite
\bea
C^{\prime 0} &=& \sqrt{2} (\frac{c_1} {N_c} + c_2 ) N_c^2 \,
{\overline{EM}}_{\sss CF}' ~~,~~~~
{\overline{EM}}_{\sss CF}' = {\overline{EM}'_{\sss C}} +
\frac{N_c}{2} \left( 1-\frac{1}{N_c^2}\right) \sum_{i=7}^{10}
{\overline{EM}}_{{\sss C}i} ~, \nn\\
C^{\prime 1} &=& \sqrt{2} \, \frac{c_1}{2} {N_c^2}\left(
1-\frac{1}{N_c^2}\right) \sum_{i=1}^6 \overline{EM}_{{\sss C}i}~.
\eea

Comparing the above expression to that of the $T'$ diagram, we have
${\overline{EM}}_{\sss F}' = {\overline{EM}}_{\sss CF}'$ in the SU(3)
limit, leading to the observation that, at $O(\alpha_s^0)$, the ratio
$C'/T'$ is independent of matrix elements [Eq.~(\ref{CTrel})].
However, if we combine the $O(\alpha_s^0)$ and $O(\alpha_s^1)$
results, the ratio is {\it not} independent of matrix elements:
\beq
{C' \over T'} = 
{(\frac{c_1}{ N_c} + c_2 ) \, {\overline{EM}}_{\sss CF}'+ \frac{c_1}{2}\left(
1-\frac{1}{N_c^2}\right) \sum_{i=1}^6 \overline{EM}_{{\sss C}i} 
\over
(c_1  + \frac{c_2}{ N_c}) \, {\overline{EM}}_{\sss F}' + \frac{c_2}{2}\left(
1-\frac{1}{N_c^2}\right) \sum_{i=1}^6 {\overline{EM}}_i } ~.
\label{fullCTrel}
\eeq
Note that the $O(\alpha_s^1)$ corrections to $C'$ and $T'$ are
proportional to $c_1$ and $c_2$, respectively. Since $c_1$ is quite a
bit larger than $c_2$, we therefore expect that the first-order
correction to the ratio $C'/T'$ is also large. However, to obtain the
exact value of $C'/T'$ beyond leading order, one needs to use a method
to estimate the matrix elements. This establishes the connection
between contractions and QCDfac/pQCD.

\section{Is there a $\btopik$ puzzle?}

The amplitudes for the four $\btopik$ decays are given in terms of
diagrams in Eq.~(\ref{ampsdiags}). Many of these diagrams are expected
to be negligible ($A'$, $\pewep$, $\pewpup$) \cite{GHLR}.  Retaining
only those which are expected to be sizeable, we have
\bea
\label{bkpiamps}
A^{+0} & = & P_{tc}' + P'_{uc} e^{i\gamma} -\frac13 \pewcp ~, \nn\\
\sqrt{2} A^{0+} & = & -T' e^{i\gamma} -C' e^{i\gamma} -P_{tc}' -
P'_{uc} e^{i\gamma} -~\pewp -\frac23 \pewcp ~, \nn\\
A^{-+} & = & -T' e^{i\gamma} - P_{tc}' -P'_{uc} e^{i\gamma} -\frac23
\pewcp ~, \nn\\
\sqrt{2}A^{00} & = & -C' e^{i\gamma} + P_{tc}' +P'_{uc} e^{i\gamma} -
\pewp -\frac13 \pewcp ~.
\eea
In the above, $P'_{tc} \equiv P'_t - P'_c$, $P'_{uc} \equiv P'_u -
P'_c$, and we have explicitly written the relative weak phase $\gamma$
(the phase information in the CKM quark mixing matrix is
conventionally parametrized in terms of the unitarity triangle, in
which the interior (CP-violating) angles are known as $\alpha$,
$\beta$ and $\gamma$ \cite{pdg}).

The diagrams $\pewp$ and $\pewcp$ are not independent -- as has been
shown, they are related to $T'$ and $C'$. If we do not make the
approximation that $c_1/c_2 = c_9/c_{10}$, we have
\bea
\label{fullEWPtreerels}
\pewp & \!\!=\!\! & {3\over 4} {c_9 + c_{10} \over c_1 + c_2} R (T' +
C') \!+\!  {3\over 4} {c_9 - c_{10} \over c_1 - c_2} R (T' - C')
~, \nn\\
\pewcp & \!\!=\!\! & {3\over 4} {c_9 + c_{10} \over c_1 + c_2} R (T' +
C') \!-\!  {3\over 4} {c_9 - c_{10} \over c_1 - c_2} R (T' - C')
~,
\eea
where 
\beq
R \equiv \left\vert {V_{tb}^* V_{ts} \over V_{ub}^* V_{us}}
\right\vert = {1 \over \lambda^2} {\sin(\beta +\gamma) \over
\sin\beta}~.
\eeq
(If $c_1/c_2 = c_9/c_{10}$ is assumed, the ratios of Eq.~(\ref{eq2})
are reproduced.)

The fact that $R$ can be expressed in terms of $\beta$ and $\gamma$
was used in Ref.~\cite{IPLL} to extract these CP phases from $\btopik$
decays. However, $\beta$ and $\gamma$ can also be obtained
independently: $\beta$ can be taken from the measurement of
mixing-induced CP violation in $\bd(t) \to J/\psi\ks$: $\sin 2\beta =
0.687 \pm 0.032$ \cite{sin2beta}, while $\gamma$ can be found via a
fit to independent measurements: $\gamma = {58.6^{+6.8}_{-5.9}}^\circ$
\cite{CKMfitter}. In the fits which follow, we include these
independent determinations of the weak phases.

Given that $\pewp$ and $\pewcp$ can be related to $T'$ and $C'$, the
amplitudes in fact depend on only seven unknown theoretical
parameters: the four magnitudes $|P'_{tc}|$, $|T'|$, $|C'|$ and
$|P'_{uc}|$, and their three relative strong phases. However, there
are nine $\btopik$ measurements: the branching ratios for the four
$\btopik$ decays, the four direct CP asymmetries $A_{dir}$, and the
mixing-induced CP asymmetry $A_{mix}$ in $\bd\to \pi^0K^0$. The latest
data is shown in Table~\ref{bpiktable}.

\begin{table}
\caption{Branching ratios, direct CP asymmetries $A_{dir}$, and
mixing-induced CP asymmetry $A_{mix}$ (if applicable) for the four
$\btopik$ decay modes.}
\center
\begin{tabular}{ c c c c }
Mode & $BR(10^{-6})$ & $A_{dir}$ & $A_{mix}$ \\ \hline
$B^+ \to \pi^+ K^0$ & $24.1 \pm 1.3$ & $-0.020 \pm 0.04$ & \\
$B^+ \to \pi^0 K^+$ & $12.1 \pm 0.8$ & ~~$0.04 \pm 0.04$ & \\
$\bd \to \pi^- K^+$ & $18.9 \pm 0.7$ & $-0.115 \pm 0.018$ & \\
$\bd \to \pi^0 K^0$ & $11.5 \pm 1.0$ & $0.02 \pm 0.13$ & $0.31 \pm
0.26$ \\
\hline\hline
\end{tabular}
\label{bpiktable}
\end{table}

It is therefore possible to perform a fit. We find a good fit:
$\chi^2_{min}/d.o.f. = 3.03/2$ \cite{thanks}. However, the fit also
gives $|C'/T'| = 1.73 \pm 1.01$, whose central value is far larger
than that given by the $O(\alpha_s^0)$ result [Eq.~(\ref{CTrel})].
This is the $\btopik$ puzzle: present $\btopik$ data seem to be
inconsistent with naive SM predictions. (Above, we used the ratio
$|C'/T'|$ to illustrate this point, but other quantities can be used
as well.)  Many analyses have found this result \cite{puzzle}. Note
that, at present, the experimental errors are still quite large (and
the theoretical errors, such as SU(3) breaking, have not been
included), so that the effect of the $\btopik$ puzzle is not yet
statistically significant. As such, it can be said to offer only a
hint of a discrepancy.

The purpose of this section is to critically re-examine the question
of whether there is a $\btopik$ puzzle. Assuming that the theoretical
uncertainties are under control, and that the effect is not a
statistical fluctuation -- and these assumptions may well be wrong --
the relevant question is: is it possible that $|C'/T'| = 1.7$?

We begin with the expression for $C'/T'$, Eq.~(\ref{fullCTrel}):
\beq
{C' \over T'} = 
{(\frac{c_1}{ N_c} + c_2 ) \, {\overline{EM}}_{\sss CF}'+ \frac{c_1}{2}\left(
1-\frac{1}{N_c^2}\right) \sum_{i=1}^6 \overline{EM}_{{\sss C}i} 
\over
(c_1  + \frac{c_2}{ N_c}) \, {\overline{EM}}_{\sss F}' + \frac{c_2}{2}\left(
1-\frac{1}{N_c^2}\right) \sum_{i=1}^6 {\overline{EM}}_i } ~.
\eeq
The important question is, to what extent does the lowest-order (naive
factorization) result for $C'/T'$ [Eq.~(\ref{CTrel})] represent the
true SM value for this ratio? Note that if $|C'/T'| = 0.17 \pm 0.1$ is
included in the fit as a constraint, we obtain a very poor fit:
$\chi^2_{min}/d.o.f. = 20.6/4$, which corresponds to a deviation from
expectations of $3.6 \sigma$ \cite{thanks}. Note also that our fit
includes $|P'_{uc}|$, and hence a large value of $P'_u$ cannot lead to
a good fit to the data (this is somewhat contrary to
Ref.~\cite{gros}).

In QCDfac the value of $|C'/T'|$ at $O(\alpha_s^1)$ may be raised to
about 2-3 times the lowest-order result \cite{BBNSKpi} and thus falls
far short of the value required to fit the data.  The fact that the
first-order correction is quite large follows from the observation
that $C'$ and $T'$ get corrections proportional to $c_1$ and $c_2$,
respectively [Eq.~(\ref{fullCTrel})], but $c_1$ is about 5-6 times
larger than $c_2$. Within pQCD, one also finds that the value of
$|C'/T'|$ may be raised to 2-3 times the lowest-order result including
NLO corrections \cite{PQCDKpi}. Note that although the authors of
Ref.~\cite{PQCDKpi} give results that are consistent with the central
values of the direct CP asymmetries in the $K \pi$ system, they cannot
explain the central value of the indirect CP asymmetry in $ K^0
\pi^0$. On the other hand, our fit takes into account {\it all} $ K
\pi$ data, including the indirect CP asymmetery in $ K^0 \pi^0$. One
therefore has to be careful about the claim in Ref~\cite{PQCDKpi} that
the $K \pi$ puzzle is resolved within pQCD.

Although we do not have an all-orders result for $|C'/T'|$ to compare
with the fit, we can still make several observations. It is true that
higher-order effects, such as those from additional gluons, will
affect the lowest-order result for $|C'/T'|$ and change its value.
However, in order to produce a value that agrees with the fit, this
value must change by a factor of 10! There are then two possibilities.
One is that the true value of $|C'/T'|$ is large, which means that the
corrections are very important. In this case, there is no $\btopik$
puzzle. However, both methods of calculating matrix elements (QCDfac
and pQCD) have argued for a reasonably small value of the strong
coupling $\alpha_s$ at the scale $m_b$. Thus, the $O(\alpha_s^n)$
corrections are increasingly small. This in turn leads to a
perturbative expansion for nonleptonic $B$ decays. Now, if there is
really no $\btopik$ puzzle, this means that the $O(\alpha_s^n)$
corrections are large, which sheds much doubt on the results of both
QCDfac and pQCD. The other possibility is that the corrections are
small and the true value of $|C'/T'|$ is similar to the
$O(\alpha_s^0)$ result. In our opinion, this situation is more likely:
we have small corrections, so that the lowest-order result for
$|C'/T'|$ [Eq.~(\ref{CTrel})] is approximately correct, the results of
QCDfac and pQCD are believable, and there {\it is} a $\btopik$ puzzle.

\section{Conclusions}

We have presented a general approach to hadronic $B$ decays. The
starting point is the observation that, at the quark level, the decay
$B\to M_1 M_2$ ($M_{1,2}$ are mesons) involves the matrix elements
$\langle \bar q_1 q_2 \bar q_3 q_4 | \bar q_5 b \, \bar q_6 q_7 | \bar
q_8 b \rangle$, where $M_1 = \bar q_1 q_2$, $M_2 =\bar q_3 q_4$, and
$\bar q_5 b \, \bar q_6 q_7$ is an operator of the effective
Hamiltonian (Dirac and colour structures are omitted here).  In order
to determine all possible ways in which the final-state quarks can be
produced, it is necessary to sum over all possible Wick contractions
of the quarks for all operators. The amplitude for $B\to M_1 M_2$ can
then be expressed as a sum over these contractions. In this paper we
have examined various properties of these contractions.

There are a total of 24 possible contractions. However, due to the
fact that we can write the final state as $M_1 M_2$ or $M_2 M_1$, not
all contractions are independent. We have shown that there are in fact
only 14 independent contractions. Buras and Silvestrini (BS) obtained
these results some years ago \cite{BS}. However, our analysis goes
beyond that of BS in several ways, described below.

We have separated the independent contractions into four classes.
Using recent theoretical and experimental developments, we have shown
that certain contractions are smaller than others, and can
occasionally be neglected. This greatly simplifies the expressions for
the amplitudes in terms of contractions.

It is also possible to write the $B$-decay amplitudes in terms of
diagrams \cite{GHLR}. However, the relation between diagrams and the
effective Hamiltonian was not made clear, and there was some question
about the rigourousness of the diagrammatic approach. We have shown
that all diagrams can be simply expressed in terms of contractions,
thereby demonstrating formally that the diagrammatic method is
rigourous.

In the limit of neglecting the (small) Wilson coefficients $C_7$ and
$C_8$, and assuming flavour SU(3) symmetry, Neubert/Rosner and
Gronau/Pirjol/Yan have shown that there are relations between the
electroweak-penguin and tree diagrams.  We show that these relations
are reproduced using the contractions method.

All of the above results hold to all orders in $\alpha_s$.  That is,
the contractions method includes the exchange of any number of gluons
between the quarks. However, we have shown that it is also possible to
work order-by-order in $\alpha_s$ by including the gluons
explicitly. This is useful, as it allows one to make a connection
between the approach of contractions and the matrix-element evaluation
methods of QCD factorization (QCDfac) and perturbative QCD (pQCD). If
one works to leading order [$O(\alpha_s^0)$], we have shown that one
finds that in the SU(3) limit the ratio $C'/T'$ is independent of
matrix elements ($T'$ and $C'$ are, respectively, the colour-allowed
and colour-suppressed tree diagrams in $\btopik$ decays). In addition,
this ratio is found the be rather small: $C'/T' \approx 0.17$. If one
adds a single gluon [$O(\alpha_s^1)$], this value changes, but one
needs to evaluate matrix elements to determine its value.

Finally, we re-examine the question of whether there is a ``$\btopik$
puzzle.'' If one considers all $\btopik$ data, a good fit is obtained,
but $|C'/T'| = 1.73 \pm 1.01$ is required. This central value is far
larger than the $O(\alpha_s^0)$ result. Leaving aside the possibility
of a statistical fluctuation, which might indeed be the true
explanation, the question of whether or not there is a $\btopik$
puzzle comes down to the question of whether or not additional gluons
can change $|C'/T'|$ from 0.17 (the $O(\alpha_s^0)$ result) to 1.7.
The higher-order effects have been evaluated in QCDfac and pQCD, and
both methods find that $|C'/T'|$ is changed by at most 2-3 times the
lowest-order result, instead of the required factor of 10. This is not
surprising as both methods find that $\alpha_s$ is relatively small at
the scale $m_b$. If $|C'/T'|$ were changed by a large amount, that
would shed doubt on the smallness of $\alpha_s$, as well as all
predictions of QCDfac and pQCD. We therefore conclude that it is
likely that $|C'/T'|$ is relatively small (i.e.\ similar to the
$O(\alpha_s^0)$ result), and that there really is a $\btopik$ puzzle.

\bigskip
\noindent {\bf Acknowledgements}:
%\bigskip

We thank J. Charles, M. Gronau, J. Rosner and L. Silvestrini for
helpful communications. M.I. thanks T. Mannel and T. Feldmann for
useful physics discussions, and acknowledges the hospitality of LAPTH
in Annecy, France and the Universit\"at Siegen, Germany, where part of
this work was done. This work is financially supported by NSERC of
Canada.

%%%%%%%%%%%%%%%%%%%%% REFERENCES %%%%%%%%%%%%%%%%%%%%%%%%%%%%%%%%


\begin{thebibliography}{99}

\bibitem{BBL} See, for example, G.~Buchalla, A.~J.~Buras and
M.~E.~Lautenbacher, Rev.\ Mod.\ Phys.\ {\bf 68}, 1125 (1996).
%%CITATION = HEP-PH 9512380;%%

\bibitem{GHLR} M. Gronau, O.F. Hern\'andez, D. London, J.L. Rosner,
Phys.\ Lett.\ B {\bf 333}, 500 (1994), Phys.\ Rev.\ D {\bf 50}, 4529
(1994), Phys.\ Rev.\ D {\bf 52}, 6356 (1995). Phys.\ Rev.\ D {\bf
52}, 6374 (1995).
%%CITATION = HEP-PH 9404281;%%
%%CITATION = HEP-PH 9404283;%%
%%CITATION = HEP-PH 9504326;%%
%%CITATION = HEP-PH 9504327;%%

\bibitem{BBNS} M.~Beneke, G.~Buchalla, M.~Neubert and C.~T.~Sachrajda,
Phys.\ Rev.\ Lett.\ {\bf 83}, 1914 (1999), Nucl.\ Phys.\ B {\bf 591},
313 (2000), Nucl.\ Phys.\ B {\bf 606}, 245 (2001).
%%CITATION = HEP-PH 9905312;%%
%%CITATION = HEP-PH 0006124;%%
%%CITATION = HEP-PH 0006124;%%
%%CITATION = HEP-PH 0104110;%%

\bibitem{PQCD} Y.~Y.~Keum, H.~n.~Li and A.~I.~Sanda, Phys.\ Lett.\ B
 {\bf 504}, 6 (2001), Phys.\ Rev.\ D {\bf 63}, 054008 (2001)
%%CITATION = HEP-PH 0004004;%% 
%%CITATION = HEP-PH 0004173;%%

\bibitem{NR} M.~Neubert and J.~L.~Rosner, Phys.\ Lett.\ B {\bf 441},
403 (1998), Phys.\ Rev.\ Lett.\ {\bf 81}, 5076 (1998).
%%CITATION = HEP-PH 9808493;%%
%%CITATION = HEP-PH 9809311;%%

\bibitem{GPY} M.~Gronau, D.~Pirjol and T.~M.~Yan, Phys.\ Rev.\ D {\bf
60}, 034021 (1999) [Erratum-ibid.\ D {\bf 69}, 119901 (2004)].
%%CITATION = HEP-PH 9810482;%%

\bibitem{SCET} C.W. Bauer, D. Pirjol, I.Z.  Rothstein, and
I.W. Stewart, Phys. Rev. D {\bf 70}, 054015 (2004); C.W. Bauer,
I.Z. Rothstein, and I.W. Stewart, hep-ph/0510241.
%%CITATION = HEP-PH 0401188;%%
%%CITATION = HEP-PH 0510241;%%

\bibitem{BS} A.~J.~Buras and L.~Silvestrini, Nucl.\ Phys.\ B {\bf
 569}, 3 (2000).
%%CITATION = HEP-PH 9812392;%%

\bibitem{Datta-Lon} A.~Datta and D.~London, Phys.\ Lett.\ B {\bf 595},
453 (2004); A.~Datta, M.~Imbeault, D.~London, V.~Page, N.~Sinha and
R.~Sinha, Phys.\ Rev.\ D {\bf 71}, 096002 (2005).
%%CITATION = HEP-PH 0404130;%%
%%CITATION = HEP-PH 0406192;%%

\bibitem{charming} M.~Ciuchini, E.~Franco, G.~Martinelli and
L.~Silvestrini, Nucl.\ Phys.\ B {\bf 501}, 271 (1997); M.~Ciuchini,
E.~Franco, G.~Martinelli, M.~Pierini and L.~Silvestrini, Phys.\ Lett.\
B {\bf 515}, 33 (2001).
%%CITATION = HEP-PH 9703353;%%
%%CITATION = HEP-PH 0104126;%%

\bibitem{sin2beta} B.~Aubert {\it et al.}  [BABAR Collaboration],
Phys.\ Rev.\ Lett.\ {\bf 94}, 161803 (2005); K.~Abe {\it et al.}
[Belle Collaboration], arXiv:hep-ex/0507037.
%%CITATION = HEP-EX 0408111;%%
%%CITATION = HEP-EX 0408127;%%

\bibitem{dattalipkin} A.~Datta, H.~J.~Lipkin and P.~J.~O'Donnell,
Phys.\ Lett.\ B {\bf 529}, 93 (2002).
%%CITATION = HEP-PH 0111336;%%

\bibitem{pdg} Particle Data Group Collaboration, S.~Eidelman {\it et
al.}, Phys.\ Lett.\ B {\bf 592} (2004) 1.
%%CITATION = PHLTA,B592,1;%%

\bibitem{IPLL} M.~Imbeault, A.~L.~Lemerle, V.~Page and D.~London,
Phys.\ Rev.\ Lett.\ {\bf 92}, 081801 (2004).
%%CITATION = HEP-PH 0309061;%%

\bibitem{CKMfitter} The CKMfitter group, http://www.slac.stanford.edu
\\ \null~~~~/xorg/ckmfitter/ckm\_results\_summerEPS2005.html

\bibitem{thanks} We thank S. Baek for the numerical results of this
fit.

\bibitem{puzzle} A.~J.~Buras, R.~Fleischer, S.~Recksiegel and
F.~Schwab, Phys.\ Rev.\ Lett.\ {\bf 92}, 101804 (2004), Nucl.\ Phys.\
B {\bf 697}, 133 (2004), Acta Phys.\ Polon.\ B {\bf 36}, 2015 (2005),
arXiv:hep-ph/0411373; C.~W.~Chiang, M.~Gronau, J.~L.~Rosner and
D.~A.~Suprun, Phys.\ Rev.\ D {\bf 70}, 034020 (2004); CKMfitter Group,
J.~Charles {\it et al.}, Eur.\ Phys.\ J.\ C {\bf 41}, 1 (2005);
S.~Mishima and T.~Yoshikawa, Phys.\ Rev.\ D {\bf 70}, 094024 (2004);
Y.~L.~Wu and Y.~F.~Zhou, Phys.\ Rev.\ D {\bf 71}, 021701 (2005);
Y.~Y.~Charng and H.~n.~Li, hep-ph/0410005; X.~G.~He and
B.~H.~J.~McKellar, hep-ph/0410098; H.~Y.~Cheng, C.~K.~Chua and
A.~Soni, Phys.\ Rev.\ D {\bf 71}, 014030 (2005); S.~Baek, P.~Hamel,
D.~London, A.~Datta and D.~A.~Suprun, Phys.\ Rev.\ D {\bf 71}, 057502
(2005).
%%CITATION = HEP-PH 0312259;%%
%%CITATION = HEP-PH 0402112;%%
%%CITATION = HEP-PH 0410407;%%
%%CITATION = HEP-PH 0411373;%%
%%CITATION = HEP-PH 0404073;%%
%%CITATION = HEP-PH 0406184;%%
%%CITATION = HEP-PH 0408090;%%
%%CITATION = HEP-PH 0409221;%%
%%CITATION = HEP-PH 0410005;%%
%%CITATION = HEP-PH 0410098;%%
%%CITATION = HEP-PH 0409317;%%
%%CITATION = HEP-PH 0412086;%%

\bibitem{gros} Y.~Grossman, A.~Hocker, Z.~Ligeti and D.~Pirjol, Phys.\
Rev.\ D {\bf 72}, 094033 (2005).
%%CITATION = HEP-PH 0506228;%%

\bibitem{BBNSKpi} M. Beneke and D. Yang, Nucl. Phys. {\bf B736}, 34
(2006); M.~Beneke and M.~Neubert, Nucl.\ Phys.\ B {\bf 675}, 333
(2003).
%%CITATION = HEP-PH 0508250;%%
%%CITATION = HEP-PH 0308039;%%

\bibitem{PQCDKpi} H.~n.~Li, S.~Mishima and A.~I.~Sanda, Phys.\ Rev.\ D
{\bf 72}, 114005 (2005).
%%CITATION = HEP-PH 0508041;%%

\end{thebibliography}
\end{document}